\documentclass[runningheads,a4paper]{llncs}

\usepackage{amssymb}
\usepackage{amsmath}
\usepackage{graphicx}
\usepackage{amsmath}
\usepackage{hyperref}
\usepackage{todonotes}
\usepackage{datetime}
\usepackage{listings}
\usepackage{xcolor}
\usepackage{wrapfig}
\usepackage{subfig}
\usepackage{booktabs}
\usepackage{url}
\usepackage{float}
\usepackage{upquote}
\usepackage{placeins}

\newcommand{\ttt}[1]{\texttt{#1}}
\newcommand{\knn}{$k$-NN }
\newcommand{\knnns}{$k$-NN}

\newcommand{\LibVersion}{1.8}
\newcommand{\replocdir}{https://github.com/searchivarius/nmslib/tree/v\LibVersion/}
\newcommand{\replocfile}{https://github.com/searchivarius/nmslib/blob/v\LibVersion/}

\begin{document}

\mainmatter  

\title{Non-Metric Space Library (NMSLIB) Manual}

\titlerunning{NMSLIB Manual}

%
%
\author{Bilegsaikhan Naidan \and Leonid Boytsov \and Yury Malkov \and David Novak}
\authorrunning{Naidan et al}


%
%


\maketitle

{\begin{center}{\small \textbf{Maintainer}: Leonid Boytsov} \email{leo@boytsov.info}\end{center}}
{\begin{center}{Version \LibVersion}\end{center} 
{\begin{center}{{\today}}\end{center}}

\begin{abstract}
This document covers a library for fast similarity (\knnns) search.
It describes \textbf{only}
search methods and distances (spaces).
Details about building, installing, Python bindings can be found online: \newline \url{\replocdir}.
Even though the library contains a variety of exact metric-space access methods,
our main focus is on more \emph{generic} and \emph{approximate} search methods,
in particular, on methods for non-metric spaces.
NMSLIB is possibly the first library with a principled support for non-metric space searching.
\end{abstract}

\section{Objectives and History}

Non-Metric Space Library (NMSLIB) is an \textbf{efficient} and \textbf{extendable} cross-platform similarity search library and a toolkit for evaluation of similarity search methods.  The core-library does \textbf{not} have any third-party dependencies.

The goal of the project is to create an effective and \textbf{comprehensive} toolkit for searching in \textbf{generic metric and non-metric} spaces.
Even though the library contains a variety of metric-space access methods,
our main focus is on \emph{generic} and \emph{approximate} search methods,
in particular, on methods for non-metric spaces.
NMSLIB is possibly the first library with a principled support for non-metric space searching.

NMSLIB is an extendible library, which means that is possible to add new search methods and distance functions. NMSLIB can be used directly in C++ and Python (via Python bindings). In addition, it is also possible to build a query server, which can be used from Java (or other languages supported by Apache Thrift). Java has a native client, i.e., it works on many platforms without requiring a C++ library to be installed.

NMSLIB started as a personal project of Bilegsaikhan Naidan, who created the initial code base, the Python bindings,
and participated in earlier evaluations. 
The most successful class of methods--neighborhood/proximity graphs--is represented by the Hierarchical Navigable Small World Graph (HNSW)
due to Malkov and Yashunin \cite{Malkov2016}. 
Other most useful methods, include a modification of the Vantage-Point Tree (VP-tree) \cite{Boytsov_and_Bilegsaikhan:nips2013},
a Neighborhood APProximation index (NAPP) \cite{tellez2013succinct},
which was improved by David Novak,
as well as a vanilla uncompressed inverted file.

The current version of the manual focuses solely on the description of:
\begin{itemize}
    \item search spaces and distances (see \S~\ref{SectionSpaces});
    \item search methods (see \S~\ref{SectionMethods}).
\end{itemize}

Details about building/installing, benchmarking, extending the code, as well as basic
tuning guidelines, can be found online: \newline \url{https://github.com/nmslib/nmslib/tree/v\LibVersion/manual/README.md}.

\section{Terminology and Problem Formulation}\label{SectionProbForm}
Similarity search is an essential part of many applications,
which include, among others,  
content-based retrieval of multimedia  and statistical machine learning.
The search is carried out in a finite database of objects $\{o_i\}$,
using a search query $q$ and a dissimilarity measure (the term data point or simply a point is often
used a synonym to denote either a data object or a query).
 The dissimilarity measure is typically represented by a distance function $d(o_i, q)$. 
The ultimate goal is to answer a query by retrieving a subset of database objects sufficiently similar to the query $q$.
These objects will be called \emph{answers}.
A combination of data points and the distance function is called a \emph{search space},
or simply a \emph{space}.

Note that we use the terms \emph{distance} and the \emph{distance function} in a broader sense than
some of the textbooks:
We do not assume that the distance is a true metric distance. 
The distance function can disobey the triangle inequality and/or be even non-symmetric.

Two retrieval tasks are typically considered: a nearest neighbor and a range search. 
The nearest neighbor search aims to find the least dissimilar object,
i.e., the object at the smallest distance from the query.
Its direct generalization is the $k$-nearest neighbor search (the \knn search),
which looks for the $k$  closest objects.
Given a radius $r$, 
the range query retrieves all objects within a query ball (centered at the query object $q$) with the radius $r$,
or, formally, all the objects~$\lbrace o_i \rbrace$ such that $d(o_i, q) \le r$. 
In generic spaces, the distance is not necessarily symmetric. 
Thus, two types of queries can be considered. 
In a  \emph{left} query, the object is the left argument of the distance function,
while the query is the right argument.
In a \emph{right} query, $q$ is the first argument and the object is the second, i.e.,
the right, argument.

The queries can be answered either exactly, 
i.e., by returning a complete result set that does not contain erroneous elements, or, 
approximately, e.g., by finding only some answers.
Thus, the methods are evaluated in terms of efficiency-effectiveness trade-offs
rather than merely in terms of their efficiency.
One common effectiveness metric is recall. In the case
of the nearest neighbor search, it is computed as
an average fraction of true neighbors returned by the method with ties broken arbitrarily.

\section{Spaces}\label{SectionSpaces}
Currently we provide implementations mostly for vector spaces.
Vector-space input files can come in either regular, i.e., dense,
or sparse variant. 
A detailed list of spaces, their parameters, 
and performance characteristics is given in Table~\ref{TableSpaces}.

The mnemonic name of the space is passed to 
python bindings function \ttt{}
as well as to
the benchmarking utility see: \newline \url{https://github.com/nmslib/nmslib/tree/v\LibVersion/manual/benchmarking.md}. 
\newline
There can be more than one version of a distance function,
which have different space-performance trade-off.
In particular, for distances that require computation of logarithms 
we can achieve an order of magnitude improvement (e.g., for the GNU C++
and Clang) by pre-computing
logarithms at index time. This comes at a price of extra storage. 
In the case of Jensen-Shannon distance functions, we can pre-compute some 
of the logarithms and accurately approximate those we cannot pre-compute.
The details are explained in \S~\ref{SectionLP}-\ref{SectionBregman}.

Straightforward slow implementations of the distance functions may have the substring \ttt{slow}
in their names, while faster versions contain the substring \ttt{fast}.
Fast functions that involve approximate computations contain additionally the substring \ttt{approx}.
For non-symmetric distance function, a space may have two variants: one variant is for left
queries (the data object is the first, i.e., left, argument of the distance function 
while the query object
is the second argument) 
and another is for right queries (the data object is the second argument and the query object is the first argument).
In the latter case the name of the space ends on \ttt{rq}.
Separating spaces by query types, might not be the best approach.
Yet, it seems to be unavoidable, because, in many cases,
we need separate indices to support left and right queries \cite{Cayton2008}.
If you know a better approach, feel free, to tell us.

\subsection{Details of Distance Efficiency Evaluation}\label{SectionDistEvalDetails}
Distance computation efficiency was evaluated on a Core i7 laptop (3.4 Ghz peak frequency)
in a single-threaded mode by the utility:
\newline \href{\replocfile similarity_search/test/bench_distfunc.cc}{bench\_distfunc}.
It is measured in millions of computations per second for single-precision
floating pointer numbers (double precision computations are, of course, more costly). 
The code was compiled using the GNU compiler. 
All data sets were small enough to fit in a CPU cache, which may have resulted in slightly more optimistic
performance numbers for cheap distances such as $L_2$.

Somewhat higher efficiency numbers can be obtained by using the Intel compiler
or the Visual Studio (Clang seems to be equally efficient to the GNU compiler).
In fact, performance is much better for distances relying on ``heavy'' math functions:
slow versions of KL- and Jensen-Shannon divergences and Jensen-Shannon metrics, 
as well as for $L_p$ spaces,
where $p \not\in\{1,2,\infty\}$.

In the efficiency test, all dense vectors have 128 elements.
For all dense-vector distances except the Jensen-Shannon divergence,
their elements were generated randomly and uniformly.
For the Jensen-Shannon divergence, we first generate elements randomly,
and next we randomly select elements that are set to zero (approximately half of all elements). 
Additionally, for KL-divergences and the JS-divergence,
we normalize vector elements so that they correspond a true discrete probability distribution. 

Sparse space distances were tested using sparse vectors from two sample files in the
\href{\replocfile sample_data}{sample\_data} directory.
Sparse vectors in 
\href{\replocfile sample_data/sparse_5K.txt}{the first} 
and
\href{\replocfile sample_data/sparse_wiki_5K.txt}{the second file} on average contain
about 100 and 600 non-zero elements, respectively.

String distances were tested using \href{\replocfile sample_data/dna32_4_5K.txt}{DNA sequences} sampled from a human genome.\footnote{\url{http://hgdownload.cse.ucsc.edu/goldenPath/hg38/bigZips/}}
The length of each string was sampled from a normal distribution $\mathcal{N}(32,4)$. 

The Signature Quadratic Form Distance (SQFD) \cite{Beecks:2010,Beecks:2013} was tested 
using signatures extracted from LSVRC-2014 data set~\cite{ILSVRCarxiv14}, 
which contains 1.2 million high resolution images.
We implemented \href{\replocfile data/data\_conv/sqfd}{our own code} to extract signatures following the method of Beecks~\cite{Beecks:2013}.
For each image, we selected $10^4$ pixels randomly and
mapped them into 7-dimensional feature space:
three color, two position, and two texture dimensions.
The features were clustered by the standard $k$-means algorithm with 20 clusters.
Then, each cluster was represented by an 8-dimensional vector, which included
a 7-dimensional centroid and a cluster weight (the number of cluster points divided
by $10^4$).

\begin{table}
\caption{Description of implemented spaces\label{TableSpaces}}
\hspace{-0.6in}\begin{tabular}{l@{\hspace{2mm}}l@{\hspace{2mm}}l}
\toprule
\textbf{Space}& \textbf{Mnemonic Name \& Formula}   & \textbf{Efficiency} \\
              &                           & (million op/sec) \\
\toprule
\multicolumn{3}{c}{\textbf{Metric Spaces}}  \\
\toprule
Hamming &  \ttt{bit\_hamming}  \hspace{7.5em} $\sum_{i=1}^n |x_i-y_i|$                  &  50-400 \\
Jaccard &  \ttt{bit\_jaccard}, \ttt{jaccard\_sparse}  \hspace{0.1em} $\sum_{i=1}^n \min(x_i, y_i) / \sum_{i=1}^n \max(x_i, y_i)$         
& 500-200, 2 \\
\cmidrule(l){1-3} 
$L_1$     &  \ttt{l1}, \ttt{l1\_sparse} \hspace{6.8em}  $\sum_{i=1}^n |x_i-y_i|$        &  35, 1.6 \\
\cmidrule(l){1-3} 
$L_2$     &  \ttt{l2}, \ttt{l2\_sparse}  \hspace{6.5em}  $\sqrt{\sum_{i=1}^n |x_i-y_i|^2}$   &   30, 1.6  \\
\cmidrule(l){1-3} 
$L_{\infty}$ &  \ttt{linf}, \ttt{linf\_sparse}  \hspace{4.5em} $\max_{i=1}^n |x_i-y_i|$      &   34 , 1.6  \\
\cmidrule(l){1-3} 
$L_p$ (generic $p \ge 1$)& \ttt{lp:p=\ldots}, \ttt{lp\_sparse:p=\ldots} \hspace{0.5em} $\left(\sum_{i=1}^n |x_i-y_i|^p\right)^{1/p}$    &  0.1-3, 0.1-1.2  \\
\cmidrule(l){1-3} 
Angular distance & \ttt{angulardist}, \ttt{angulardist\_sparse}, \ttt{angulardist\_sparse\_fast} & { 13, 1.4, 3.5 } \\
                        & $\arccos\left(\frac{\sum_{i=1}^n x_i y_i}{\sqrt{\sum_{i=1}^n x_i^2}\sqrt{\sum_{i=1}^n y_i^2 }}\right)$   & \\
\cmidrule(l){1-3} 
Jensen-Shan. metr. &\ttt{jsmetrslow, jsmetrfast, jsmetrfastapprox} &  0.3, 1.9, 4.8  \\
                          & $\sqrt{\frac{1}{2}\sum_{i=1}^n \left[x_i \log x_i + y_i \log y_i  - (x_i+y_i)\log \frac{x_i +y_i}{2}\right]}$  & \vspace{1em} \\
\cmidrule(l){1-3} 
Levenshtein       &\ttt{leven} (see \S~\ref{SectionEditDistance} for details) & 0.2 \\
\cmidrule(l){1-3} 
SQFD              & \ttt{sqfd\_minus\_func}, \ttt{sqfd\_heuristic\_func:alpha=\ldots}, & 0.05, 0.05, 0.03 \\
                  &  \ttt{sqfd\_gaussian\_func:alpha=\ldots} (see \S~\ref{SectionSQFD} for details) & \\
\toprule
\multicolumn{3}{c}{\textbf{Non-metric spaces (symmetric distance)}}  \\
\toprule
$L_p$ (generic $p < 1$)& \ttt{lp:p=\ldots, lp\_sparse:p=\ldots}  \hspace{1em} $\left(\sum_{i=1}^n |x_i-y_i|^p\right)^{1/p}$   &  0.1-3, 0.1-1 \\
\cmidrule(l){1-3} 
Jensen-Shan. div. &\ttt{jsdivslow, jsdivfast, jsdivfastapprox} &   0.3, 1.9, 4.8 \\
                          & $\frac{1}{2}\sum_{i=1}^n \left[x_i \log x_i + y_i \log y_i  - (x_i+y_i)\log \frac{x_i +y_i}{2}\right]$ & \\
\cmidrule(l){1-3} 
Cosine distance & \ttt{cosinesimil}, \ttt{cosinesimil\_sparse}, \ttt{cosinesimil\_sparse\_fast} & { 13, 1.4, 3.5 } \\
                        & $1-\frac{\sum_{i=1}^n x_i y_i}{\sqrt{\sum_{i=1}^n x_i^2}\sqrt{\sum_{i=1}^n y_i^2 }}$   & \vspace{1em} \\
\cmidrule(l){1-3} 
Norm. Levenshtein       &\ttt{normleven}, see \S~\ref{SectionEditDistance} for details & 0.2 \\
\toprule
\multicolumn{3}{c}{\textbf{Non-metric spaces (non-symmetric distance)}}  \\
\toprule
Regular KL-div. & left queries: \ttt{kldivfast}       & 0.5, 27 \\
                       & right queries: \ttt{kldivfastrq}    &  \\
                       & $\sum_{i=1}^n   x_i \log \frac{x_i}{y_i}$  & \\ 
\cmidrule(l){1-3} 
Generalized KL-div. & left queries: \ttt{kldivgenslow}, \ttt{kldivgenfast} & 0.5, 27    \\
                           & right queries: \ttt{kldivgenfastrq} & 27    \\
                           & $\sum_{i=1}^n \left[  x_i \log \frac{x_i}{y_i} -   x_i +   y_i \right]$   &   \\
\cmidrule(l){1-3} 
Itakura-Saito & left queries: \ttt{itakurasaitoslow, itakurasaitofast}   & 0.2, 3, 14 \\
              & right queries: \ttt{itakurasaitofastrq}                  & 14         \\
              & $\sum_{i=1}^n \left[ \frac{ x_i}{y_i} - \log \frac{x_i}{y_i}  -1 \right]$ \\
R\'{e}nyi divergence &   \ttt{renyidiv\_slow}, \ttt{renyidiv\_fast} \vspace{3em} $\frac{1}{\alpha-1}\log\left[\sum\limits_{i=1}^m x_i^\alpha y_i^{1-\alpha}\right]$      
& 0.4, 0.5-1.5
\\ 
\toprule
\end{tabular}
\end{table}

\subsection{$L_p$-norms and the Hamming Distance}\label{SectionLP}
The $L_p$ distance between vectors $x$ and $y$ are
given by the formula:
\begin{equation}\label{EqMink}
L_p(x,y) = \left(\sum_{i=1}^n |x_i-y_i|^p\right)^{1/p}
\end{equation}
In the limit ($p \rightarrow \infty$),
the $L_p$ distance becomes the Maximum metric, also known as 
the Chebyshev distance:
\begin{equation}\label{EqCheb}
L_{\infty}(x,y) = \max\limits_{i=1}^n |x_i-y_i|
\end{equation}
$L_{\infty}$ and all spaces $L_p$ for $p \ge 1$
are true metrics. 
They are symmetric, equal to zero only for identical elements,
and, most importantly, satisfy \emph{the triangle inequality}.
However, the $L_p$ norm is \emph{not} a metric if $p<1$.

In the case of dense vectors, 
we have reasonably efficient implementations 
for $L_p$ distances where $p$ is either integer or infinity. 
The most efficient implementations are for $L_1$ (Manhattan),
$L_2$ (Euclidean), and $L_{\infty}$ (Chebyshev).
As explained in the author's blog,
\href{http://searchivarius.org/blog/efficient-exponentiation-square-rooting}{we compute exponents through square rooting}. 
This works best when the number of digits (after the binary digit) is small, e.g., if $p=0.125$.

Any $L_p$ space can have a dense and a sparse variant.
Sparse vector spaces have their own mnemonic names, which are different
from dense-space mnemonic names in that they contain a suffix \ttt{\_sparse} (see also Table~\ref{TableSpaces}).
For instance \ttt{l1} and \ttt{l1\_sparse} are both $L_1$ spaces,
but the former is dense and the latter is sparse.
The mnemonic names of $L_1$, $L_2$, and $L_\infty$ spaces (passed to the benchmarking utility) are
\ttt{l1}, \ttt{l2}, and \ttt{linf}, respectively.
Other generic $L_p$ have the name \ttt{lp}, which is used in combination with a parameter.
For instance, $L_3$ is denoted as \ttt{lp:p=3}.

Distance functions for sparse-vector spaces are far less efficient, 
due to a costly, branch-heavy, operation of matching sparse vector indices
(between two sparse vectors).

In the special case of $L_1$ for binary vectors, the $L_1$ distance becomes the Hamming distance.
This case is represented by the \ttt{bit\_hamming} space, where data points are stored as compact
bit vectors.

\subsection{Scalar-product Related Distances}
We have two distance function whose formulas include normalized scalar product.
One is the cosine distance, which is equal to:
$$
d(x,y) =1-\frac{\sum_{i=1}^n x_i y_i} 
{\sqrt{\sum_{i=1}^n x_i^2} \sqrt{\sum_{i=1}^n y_i^2 } } 
$$ 
The cosine distance is not a true metric, but it can be converted into
one by applying a monotonic transformation (i.e.., subtracting the 
cosine distance from one and taking an inverse cosine).
The resulting distance function is a true metric, which is called the angular distance.
The angular distance is computed using the following formula:
$$
d(x,y) =\arccos\left(\frac{\sum_{i=1}^n x_i y_i} 
{\sqrt{\sum_{i=1}^n x_i^2} \sqrt{\sum_{i=1}^n y_i^2 } }\right) 
$$ 

In the case of sparse spaces, to compute the scalar product,
we need to obtain an intersection of vector element ids
corresponding to non-zero elements.
A classic text-book intersection algorithm (akin to a merge-sort)
is not particularly efficient, apparently,
due to frequent branching.
For \emph{single-precision} floating point vector elements,
we provide a more efficient implementation that relies on the 
all-against-all comparison SIMD instruction \texttt{\_mm\_cmpistrm}.
This implementation  (inspired by the set intersection algorithm of Schlegel~et~al.~\cite{schlegel2011fast})
is about 2.5-3 times faster than a pure C++ implementation based on the merge-sort approach.

\subsection{Jaccard Distance}\label{SectionJaccard}
The Jaccard distance is true metric. 
Given two binary vectors $x$ and $y$,
the Jaccard distance (also called the index) is computed using the following formula:
$$
\frac{\sum_{i=1}^n \min(x_i, y_i)}{\sum_{i=1}^n \max(x_i, y_i)}
$$

For the Jaccard distance, there are two ways to represent data:
\begin{itemize}
    \item A sparse set of dimensions (space \ttt{jaccard\_sparse});
    \item A dense binary bit vector (space \ttt{bit\_jaccard}).
\end{itemize}

\subsection{Jensen-Shannon Divergence}\label{SectionJS}
\emph{Jensen-Shannon} divergence is a symmetrized and smoothed KL-divergence:
\begin{equation}\label{EqJS}
\frac{1}{2}\sum_{i=1}^n\left[ x_i \log x_i + y_i \log y_i  -(x_i+y_i)\log \frac{x_i +y_i}{2}\right]
\end{equation}
This divergence is symmetric, but it is not a metric function.
However, the square root of the Jensen-Shannon divergence
is a proper a metric \cite{endres2003new},
which we call the Jensen-Shannon metric.

A straightforward implementation of Eq.~\ref{EqJS} is inefficient for two reasons 
(at least when one uses the GNU C++ compiler)
(1) computation of logarithms is a slow operation (2)
the case of zero $x_i$ and/or $y_i$ requires conditional processing, i.e.,
costly branches.

A better method is to pre-compute logarithms of data at index time. 
It is also necessary to compute logarithms of a query vector.
However, this operation has a little cost since it is carried out once 
for each nearest neighbor or range query.
Pre-computation leads to a 3-10 fold improvement depending on the sparsity of vectors,
albeit at the expense of requiring twice as much space.
Unfortunately, it is not possible to avoid computation of the third logarithm:
it needs to be computed in points that are not known until we see the query vector.

However, it is possible to approximate it with a very good precision,
which should be sufficient for the purpose of approximate searching.
Let us rewrite Equation \ref{EqJS} as follows:
$$
\frac{1}{2}\sum_{i=1}^n\left[ x_i \log x_i + y_i \log y_i  -(x_i+y_i)\log \frac{x_i +y_i}{2}\right]=
$$
$$
 = \frac{1}{2}\sum_{i=1}^n\left[ x_i \log x_i + y_i \log y_i\right]  -
\sum_{i=1}^n\left[\frac{(x_i+y_i)}{2}\log \frac{x_i +y_i}{2} \right]=
$$
$$
 = \frac{1}{2}\sum_{i=1}^n x_i \log x_i + y_i \log y_i  -
$$
\begin{equation}\label{Eq1}
\sum_{i=1}^n\frac{(x_i+y_i)}{2}\left[\log\frac{1}{2} + \log \max(x_i,y_i) + 
\log \left(1 + \frac{\min(x_i,y_i)}{\max(x_i,y_i)}\right) \right]
\end{equation}
We can pre-compute all the logarithms in Eq.~\ref{Eq1} except for $\log \left(1 + \frac{\min(x_i,y_i)}{\max(x_i,y_i)}\right) $. However, its argument value is in a small range: from one to two.
We can discretize the range, compute logarithms in many intermediate points and save the computed values in a table.
Finally, we employ the SIMD instructions to implement this approach. 
This is a very efficient approach, which results in a very little (around $10^{-6}$ on average) relative error for the value of the Jensen-Shannon divergence.

Another possible approach is to use \href{http://fastapprox.googlecode.com/svn/trunk/fastapprox/src/fastonebigheader.h}{an efficient approximation for logarithm computation}.
\href{https://github.com/searchivarius/BlogCode/tree/master/2013/12/26}{As our tests show},
this method is about 1.5x times faster (1.5 vs 1.0 billions of logarithms per second),
but for the logarithms in the range $[1,2]$,
 the relative error is one order magnitude higher (for a single logarithm) than for the table-based discretization approach.

\subsection{Bregman Divergences}\label{SectionBregman}
Bregman divergences are typically non-metric distance functions,
which are equal to a difference between some convex differentiable function $f$
and its first-order Taylor expansion \cite{Bregman:1967,Cayton2008}. 
More formally, given the convex and differentiable function $f$
(of many variables), its
corresponding Bregman divergence $d_f(x,y)$ is equal to:
$$
d_f(x,y) = f(x) - f(y) - \left( f(y) \cdot ( x - y ) \right)
$$
where $ x \cdot y$ denotes the scalar product of vectors $x$ and $y$.
In this library, we implement the generalized KL-divergence 
and the Itakura-Saito divergence,
which correspond to functions $f=\sum x_i \log x_i - \sum x_i$ and $f = - \sum \log x_i$. 
The generalized KL-divergence is equal to:
$$
\sum_{i=1}^n \left[  x_i \log \frac{x_i}{y_i} -   x_i +   y_i \right],
$$
while the Itakura-Saito divergence is equal to:
$$ 
\sum_{i=1}^n \left[ \frac{ x_i}{y_i} - \log \frac{x_i}{y_i}  -1 \right].
$$
If vectors $x$ and $y$ are proper probability distributions, $\sum x_i = \sum y_i = 1$.
In this case, the generalized KL-divergence becomes a regular KL-divergence:
$$
\sum_{i=1}^n \left[  x_i \log \frac{x_i}{y_i} \right].
$$

Computing logarithms is costly: We can considerably improve efficiency of 
Itakura-Saito divergence and KL-divergence by pre-computing logarithms at index time.
The spaces that implement this functionality contain the substring \ttt{fast} in their mnemonic names (see also Table~\ref{TableSpaces}).

\subsection{R\'{e}nyi Divergence}\label{SectionRenyiDiv}
The R\'{e}nyi divergence is a family of generally non-symmetric distances computed by the formula:
$$
 \frac{1}{\alpha-1}\log\left[\sum\limits_{i=1}^m x_i^\alpha y_i^{1-\alpha}\right]
$$
The value of the parameter $\alpha$ should be greater than zero. For all values
except $\alpha = 0.5$ these distances are non-symmetric.
There are two variants of each space: \ttt{renyidiv\_slow}
and \ttt{renyidiv\_fast}.
They have a parameter \ttt{alpha}.
The slower variant computes the exponents using a standard function \ttt{pow}.
For the fast variant, as explained in the author's blog,
\href{http://searchivarius.org/blog/efficient-exponentiation-square-rooting}{we compute exponents through square rooting}. 
This works best when the number of digits (after the binary digit) is small, e.g., if $p=0.125$.
The slow variant can actually be faster, if the compiler, e.g., MS Visual Studio or the Intel Compiler,
implements an efficient approximate variant of the \ttt{pow} function.

\subsection{String Distances}\label{SectionEditDistance}
We currently provide implementations for the Levenshtein distance and its length-normalized variant.
The \emph{original} Levenshtein distance is equal to the minimum number of insertions, deletions, and
substitutions (but not transpositions) required to obtain one string from another \cite{Levenshtein:1966}.
The distance between strings $p$ and $s$ is computed using the classic $O(m \times n)$
dynamic programming solution, where $m$ and $n$ are lengths of strings $p$ and $s$, respectively. 
The \emph{normalized} Levenshtein distance is obtained by dividing the original Levenshtein distance by
the maximum of string lengths. If both strings are empty, the distance is equal to zero.

While the original Levenshtein distance is a metric distance, the normalized Levenshtein function is not,
because the triangle inequality may not hold.
In practice, when there is little variance in string length, 
the violation of the triangle inequality is infrequent and, thus, the normalized Levenshtein distance
is approximately metric for many real data sets.

Technically, the classic Levenshtein distance is equal to $C_{n,m}$, where $C_{i,j}$ is computed via the classic recursion:
\begin{equation} \label{EqLeven}
\hspace{-5em}C_{i,j} =\min \left\{\begin{array}{ll} 
0 ,                 & \mbox{ if } i =  j = 0 \\
{C_{i - 1, j} +1,}  & \mbox{ if } i> 0 \\
{C_{i,j-1} +1 ,  }  & \mbox{ if } j > 0 \\
{C_{i - 1, j - 1} + \left[p_{i} \neq s_{j} \right]} ,   & \mbox{ if } i,j > 0  \\
\end{array}
\right.
\end{equation}
Because computation time is proportional to both strings' length,
this can be a costly operation: for the sample data set described in \S~\ref{SectionDistEvalDetails}, 
it is possible to compute only about 200K distances per second.

The classic algorithm to compute the Levenshtein distance was independently discovered by several researchers in
various contexts, including speech recognition \cite{Vintsyuk:1968,Velichko_and_Zagoruyko:1970,Sakoe_and_Chiba:1971} and computational
biology \cite{Needleman_and_Wunsch:1970}
 (see Sankoff \cite{Sankoff:2000} for a historical perspective).
Despite the early discovery, the algorithm was generally unknown
before a publication by Wagner and Fischer \cite{Wagner_and_Fischer:1974} in a computer science journal.

\subsection{Signature Quadratic Form Distance (SQFD)}\label{SectionSQFD}
Images can be compared using a \emph{family} of metric functions called  the
Signature Quadratic Form Distance (SQFD).
During the preprocessing stage, each image is converted to a set of $n$ signatures (the number of signatures $n$ is a parameter).
To this end, a fixed number of pixels is randomly selected.
Then, each pixel is represented by a 7-dimensional vector with the following components:
three color, two position, and two texture elements.
These 7-dimensional vectors are clustered by the standard $k$-means algorithm with $n$ centers.
Finally, each cluster is represented by an 8-dimensional vector, called \emph{signature}.
A signature includes a 7-dimensional centroid and a cluster weight (the number
of cluster points divided by the total number of randomly selected pixels).
Cluster weights form a \emph{signature histogram}.
 
The SQFD is computed as a quadratic form applied to a $2n$-dimensional vector constructed
by combining images' signature histograms. 
The combination vector includes
$n$ unmodified signature histogram values of the first image 
followed by $n$ negated signature histogram values of the second image. 
Unlike the classic quadratic form distance, where the quadratic form matrix is fixed,
in the case of the SQFD, the matrix is re-computed for each pair of images.
This can be seen as computing the distance between infinite-dimensional vectors
each of which has only a finite number of non-zero elements.

To compute the quadratic form matrix, we introduce the new global enumeration of signatures,
in which a signature $k$ from the first image has number $k$, while
the signature $k$ from the second image has number $n+k$.
To obtain a quadratic form matrix element in row $i$ column $j$ 
we first compute the Euclidean distance $d$ between the $i$-th and the $j$-th signature.
Then, the value $d$ is transformed using one of the three functions:
negation (the \emph{minus} function $-d$), a \emph{heuristic} function $\frac{1}{\alpha + d}$,
and the \emph{Gaussian} function $\exp(-\alpha d^2)$.
The larger is the distance, the smaller is the coefficient in the matrix of the quadratic form.

Note that the SQFD is a family of distances parameterized by the choice of the transformation
function and $\alpha$.
For further details, please, see the thesis of Beecks~\cite{Beecks:2013}.


\section{Search Methods}\label{SectionMethods}
Implemented search methods can be broadly divided into the following 
categories:
\begin{itemize}
\item Space partitioning methods (including a specialized method bbtree for Bregman divergences) \S~\ref{SectionSpacePartMeth};
\item Locality Sensitive Hashing (LSH) methods \S~\ref{SectionLSH};
\item Filter-and-refine methods based on projection to a lower-dimensional space \S~\ref{SectionProjMethod};
\item Filtering methods based on permutations \S~\ref{SectionPermMethod};
\item Methods that construct a proximity graph \S~\ref{SectionProxGraph};
\item Miscellaneous methods \S~\ref{SectionMiscMeth}.
\end{itemize}

In the following subsections (\S~\ref{SectionSpacePartMeth}-\ref{SectionMiscMeth}),
we describe implemented methods, explain their parameters,
and provide examples of their use via the benchmarking utility \ttt{experiment} (\ttt{experiment.exe} on Windows).
The details on building the utility \ttt{experiment} can be found online: \newline \url{https://github.com/nmslib/nmslib/tree/v\LibVersion/manual/README.md}.

\subsection{Space Partitioning Methods} \label{SectionSpacePartMeth} 
Parameters of space partitioning methods are summarized in Table~\ref{TableSpaceMethPart}.
Most of these methods are hierarchical partitioning methods.

Hierarchical space partitioning methods create a hierarchical decomposition of the space
(often in a recursive fashion),
which is best represented by a tree (or a forest). 
There are two main partitioning approaches: 
pivoting and compact partitioning schemes  \cite{Chavez_et_al:2001a}.

Pivoting methods rely on embedding into a vector space where vector elements are distances
from the object to pivots.
Partitioning is based on how far (or close) the data points are located with respect to pivots.
\footnote{If the original space is metric, mapping an object to a vector of distances to pivots 
defines the contractive embedding in the metric spaces with $L_{\infty}$ distance.
That is, the $L_{\infty}$ distance in the target vector space is a lower bound for the original distance.}

Hierarchical partitions produced by pivoting methods lack locality: a single partition can contain
not-so-close data points. In contrast, compact partitioning schemes exploit locality.
They either divide the data into clusters or create, possibly approximate, Voronoi partitions.
In the latter case, for example, we can select several centers/pivots $\pi_i$ and associate 
data points with the closest center.

If the current partition contains fewer than \ttt{bucketSize} (a method parameter) elements,
we stop partitioning of the space and place all elements
belonging to the current partition into a single bucket. 
If, in addition, the value of the parameter \ttt{chunkBucket} is set to one,
we allocate a new chunk of memory that contains a copy of all bucket vectors.
This method often halves retrieval time
at the expense of extra memory consumed by a testing utility (e.g., \ttt{experiment}) as it does not deallocate memory occupied by the original vectors. \footnote{Keeping original vectors simplifies the testing workflow.
However, this is not necessary for a real production system.
Hence, storing bucket vectors at contiguous memory locations does not have to result in a larger memory footprint.}

Classic hierarchical space partitioning methods for metric spaces are exact. 
It is possible to make them approximate via an early termination technique,
where we terminate the search after exploring a pre-specified number of partitions.
To implement this strategy, we define an order of visiting partitions.
In the case of clustering methods, we first visit partitions that are closer to a query point.
In the case of hierarchical space partitioning methods such as the VP-tree,
we greedily explore partitions containing the query.

In NMSLIB, the early termination condition is defined in terms of 
the maximum number of buckets (parameter \ttt{maxLeavesToVisit})
to visit before terminating the search procedure.
By default, the parameter \ttt{maxLeavesToVisit} is set to a large number (2147483647), 
which means that no early termination is employed.
The parameter \ttt{maxLeavesToVisit} is supported by many, but not all
space partitioning methods.

\subsubsection{VP-tree}\label{SectionVPtree}
A VP-tree \cite{Uhlmann:1991,Yianilos:1993} (also known as a ball-tree)
is a pivoting method.
During indexing, a (random) pivot is selected and a set of data objects is divided into
two parts based on the distance to the pivot.
If the distance is smaller than the median distance, the objects are placed into
one (inner) partition. If the distance is larger than the median,
the objects are placed into the other (outer) partition.
If the distance is exactly equal to the median, the placement can be arbitrary.

We attempt to select the pivot multiple times. Each time, we measure the variance of distances to the pivot.
Eventually, we use the pivot that corresponds to the maximum variance.
The number of attempts to select the pivot is controlled by the index-time parameter \ttt{selectPivotAttempts}.

The VP-tree in metric spaces is an exact search method, which relies on the triangle inequality.
It can be made approximate by applying the early termination strategy (as described 
in the previous subsection).
Another approximate-search approach, 
which is currently implemented only for the VP-tree, 
is based on the relaxed version of the triangle inequality.

Assume that $\pi$ is the pivot in the VP-tree, $q$ is the query with the radius $r$, 
and $R$ is the median distance from $\pi$ to every other data point.
Due to the triangle inequality, pruning is possible only if $r \le |R - d(\pi, q)|$. 
If this latter condition is true, 
we visit only one partition that contains the query point.
If $r > |R - d(\pi, q)|$, there is no guarantee that all answers
are in the same partition as $q$. Thus, to guarantee retrieval of all answers,
we need to visit both partitions.

The pruning condition based on the triangle inequality can be overly pessimistic.
By selecting some $\alpha > 1$ and opting to prune when $r \le \alpha |R - d(\pi, q)|$,
we can improve search performance at the expense of missing some valid answers.
The efficiency-effectiveness trade-off is affected by the choice of $\alpha$:
Note that for some (especially low-dimensional) data sets, a modest
loss in recall (e.g., by 1-5\%) can lead to an order of magnitude faster retrieval.
Not only the triangle inequality can be overly pessimistic in metric spaces, 
but it often fails to capture the geometry of non-metric spaces.
As a result, if the metric space method is applied to a non-metric space,
the recall can be too low or retrieval time can be too long.

Yet, in non-metric spaces, 
it is often possible to answer queries, 
when using $\alpha$ possibly smaller than one \cite{Boytsov_and_Bilegsaikhan:nips2013,naidan2015permutation}.
More generally, we assume that there exists an unknown decision/pruning function $D(R, d(\pi, q))$ and
that pruning is done when $r \le D(R, d(\pi, q))$.
The decision function $D()$, which can be learned from data, is called a search oracle.
A pruning algorithm based on the triangle inequality 
is a special case of the search oracle described by the formula:
\begin{equation}\label{EqDecFunc}
D_{\pi,R}(x) = \left\{
\begin{array}{ll}
\alpha_{left} |x - R|^{{exp}_{left}},  & \mbox{ if }x \le R\\
\alpha_{right} |x - R|^{{exp}_{right}}, & \mbox{ if }x \ge R\\
\end{array}
\right.
\end{equation}
There are several ways to obtain/specify optimal parameters for the VP-tree: 
\begin{itemize} 
\item using the auto-tuning procedure fired before creation of the index; 
\item using the standalone tuning utility \texttt{tune\_vptree} (\texttt{tune\_vptree.exe} for Windows);
\item fully manually.
\end{itemize}

It is, perhaps, easiest to initiate the tunning procedure during creation of the index.
To this end, one needs to specify parameters \texttt{desiredRecall} (the minimum desired recall), \texttt{bucketSize} (the size 
of the bucket), \texttt{tuneK} or \texttt{tuneR}, and (optionally) parameters \texttt{tuneQty}, \ttt{minExp} and \ttt{maxExp}. 
Parameters \texttt{tuneK} and \texttt{tuneR} are used to specify the value of $k$ for \knn search,
or the search radius $r$ for the range search.

The parameter \texttt{tuneQty} defines the maximum number of records in a subset that is used for tuning. 
The tunning procedure will sample \texttt{tuneQty} records from the main set to make a (potentially) smaller data test. 
Additional query sets will be created by further random sampling of points from this smaller data set.

The tuning procedure considers all possible values for exponents between \ttt{minExp} and \ttt{maxExp} with 
a restriction that  $exp_{left}=exp_{right}$. 
By default, \ttt{minExp} = \ttt{maxExp} = 1, which is usually a good setting.
For each value of the exponents,
the tunning procedure carries out a grid-like search procedure for parameters $\alpha_{left}$ and $\alpha_{right}$ 
with several random restarts. 
It creates several indices for the tuning subset and runs a batch of mini-experiments to
find parameters yielding the desired recall value at the minimum cost.
If it is necessary to produce more accurate estimates, the tunning method may use automatically adjusted
values for parameters \texttt{tuneQty}, \texttt{bucketSize}, and \texttt{desiredRecall}.
The tunning algorithm cannot adjust the parameter \texttt{maxLeavesToVisit}:
please, do not use it with the auto-tunning procedure.

The disadvantage of automatic tuning is that it might fail to obtain
parameters for a desired recall level. Another limitations is that a tunning procedure cannot
run on very small data sets (less than two thousand entries).

The standalone tuning utility \texttt{tune\_vptree} exploits an almost identical tuning procedure.
It differs from index-time auto-tuning in several ways:  
\begin{itemize}
\item  It can be used with other VP-tree based methods, 
in particular, with the projection VP-tree (see \S~\ref{SectionProjVPTree}).
\item It allows the user to specify a separate query set, which can be useful
when queries cannot be accurately modelled by a bootstrapping approach (sampling queries from the main data set).
\item Once the optimal values are computed, they can be further re-used without the need
to start the tunning procedure each time the index is created.
\item However, the user is fully responsible for specifying the size of the test data set
and the value of the parameter \texttt{desiredRecall}: the system
will not try to change them for optimization purposes.
\end{itemize}

If automatic tunning fails,
the user can restart the procedure with the smaller value of \ttt{desiredRecall}.
Alternatively, the user can manually specify values of  parameters:
\ttt{alphaLeft}, \ttt{alphaRight}, \ttt{expLeft}, and \ttt{expRight} (by default exponents are one).

The following is an example of testing the VP-tree with the benchmarking utility \ttt{experiment}
without the auto-tunning (note the separation into index- and query-time parameters):
{
\footnotesize
\begin{verbatim}
release/experiment \
  --distType float --spaceType l2 --testSetQty 5 --maxNumQuery 100 \
  --knn 1 --range 0.1 \
  --dataFile ../sample_data/final8_10K.txt --outFilePrefix result \
  --method vptree \
    --createIndex  bucketSize=10,chunkBucket=1 \
    --queryTimeParams alphaLeft=2.0,alphaRight=2.0,\
                      expLeft=1,expRight=1,\
                      maxLeavesToVisit=500
                 
\end{verbatim}
}

To initiate auto-tuning, one may use the following command line (note that we do 
not use the parameter \texttt{maxLeavesToVisit} here):
{
\footnotesize
\begin{verbatim}
release/experiment \
  --distType float --spaceType l2 --testSetQty 5 --maxNumQuery 100 \
  --knn 1 --range 0.1 \
  --dataFile ../sample_data/final8_10K.txt --outFilePrefix result \
  --method vptree \
    --createIndex tuneK=1,desiredRecall=0.9,\
                  bucketSize=10,chunkBucket=1
\end{verbatim}
}

\subsubsection{Multi-Vantage Point Tree}
It is possible to have more than one pivot per tree level.
In the binary version of the multi-vantage point tree (MVP-tree),
which is implemented in NMSLIB,
there are two pivots.
Thus, each partition divides the space into four parts,
which are similar to partitions created by two levels of the VP-tree.
The difference is that the VP-tree employs three pivots
to divide the space into four parts,
while in the MVP-tree two pivots are used.

In addition, in the MVP-tree we memorize 
distances between a data object and the first \ttt{maxPathLen} (method parameter)
pivots on the path connecting the root and the leaf that stores this data object.
Because mapping an object to a vector of distances (to \ttt{maxPathLen} pivots)
defines the contractive embedding in the metric spaces with $L_{\infty}$ distance,
these values can be used to improve the filtering capacity of the MVP-tree
and, consequently to reduce the number of distance computations.

The following is an example of testing the MVP-tree with the benchmarking utility \ttt{experiment}:
{
\footnotesize
\begin{verbatim}
release/experiment \
  --distType float --spaceType l2 --testSetQty 5 --maxNumQuery 100 \
  --knn 1 --range 0.1 \
  --dataFile ../sample_data/final8_10K.txt --outFilePrefix result \
  --method mvptree \
    --createIndex maxPathLen=4,bucketSize=10,chunkBucket=1 \
    --queryTimeParams maxLeavesToVisit=500
\end{verbatim}
}

Our implementation of the MVP-tree permits to answer queries both exactly and approximately (by specifying
the parameter \ttt{maxLeavesToVisit}). Yet, this implementation should be used only with metric spaces.

\begin{table}
\caption{Parameters of space partitioning methods\label{TableSpaceMethPart}}
\centering
\begin{tabular}{l@{\hspace{2mm}}p{3.5in}}
\toprule
\multicolumn{2}{c}{\textbf{Common parameters}}\\
\cmidrule(l){1-2} 
\ttt{bucketSize}    & A maximum number of elements in a bucket/leaf.    \\
\ttt{chunkBucket}   & Indicates if bucket elements should be stored contiguously in memory (1 by default).  \\
\ttt{maxLeavesToVisit}  & An early termination parameter equal to the maximum number of buckets (tree leaves) visited by a search algorithm (2147483647 by default). \\
\cmidrule(l){1-2} 
\multicolumn{2}{c}{\textbf{VP-tree} (\ttt{vptree}) \cite{Uhlmann:1991,Yianilos:1993}  } 
\\
\cmidrule(l){1-2} 
                   & Common parameters: \ttt{bucketSize}, \ttt{chunkBucket}, and \ttt{maxLeavesToVisit} \\
 \ttt{selectPivotAttempts} & A number of pivot selection attempts (5 by default) \\
 \ttt{alphaLeft}/\ttt{alphaRight}   & A stretching coefficient $\alpha_{left}$/$\alpha_{right}$ in Eq.~(\ref{EqDecFunc}) \\
 \ttt{expLeft}/\ttt{expRight} & The left/right exponent in Eq.~(\ref{EqDecFunc}) \\
 \ttt{tuneK}       & The value of $k$ used in the auto-tunning procedure (in the case of \knn search) \\
 \ttt{tuneR}       & The value of the radius $r$ used in the auto-tunning procedure (in the case of the range search) \\
 \ttt{minExp}/\ttt{maxExp} & The minimum/maximum value of exponent used in the auto-tunning procedure \\
\cmidrule(l){1-2} 
\multicolumn{2}{c}{\textbf{Multi-Vantage Point Tree} (\ttt{mvptree})  \cite{bozkaya1999indexing}}   \\
\cmidrule(l){1-2} 
                   & Common parameters: \ttt{bucketSize}, \ttt{chunkBucket}, and \ttt{maxLeavesToVisit} \\
 \ttt{maxPathLen}  & the maximum number of top-level pivots for which we memorize distances
to data objects in the leaves \\
\cmidrule(l){1-2} 
\multicolumn{2}{c}{\textbf{GH-tree} (\ttt{ghtree})  \cite{Uhlmann:1991}}   \\
\cmidrule(l){1-2} 
                   & Common parameters: \ttt{bucketSize}, \ttt{chunkBucket}, and \ttt{maxLeavesToVisit} \\
\cmidrule(l){1-2} 
\multicolumn{2}{c}{\textbf{List of clusters} (\ttt{list\_clusters})  \cite{chavez2005compact}}   \\
\cmidrule(l){1-2} 
                   & Common parameters: \ttt{bucketSize}, \ttt{chunkBucket}, and \ttt{maxLeavesToVisit}. Note \ttt{maxLeavesToVisit} is a \textbf{query-time} parameter. \\
\ttt{useBucketSize} & If equal to one, we use the parameter \ttt{bucketSize} to determine the number of points in the cluster. Otherwise, the size of the cluster is defined by the parameter \ttt{radius}. \\
\ttt{radius}        & The maximum radius of a cluster (used when \ttt{useBucketSize} is set to zero). \\
\ttt{strategy}      & A cluster selection strategy. It is one of the following: \ttt{random}, \ttt{closestPrevCenter}, \ttt{farthestPrevCenter}, \ttt{minSumDistPrevCenters}, \ttt{maxSumDistPrevCenters}. \\ 
\cmidrule(l){1-2} 
\multicolumn{2}{c}{\textbf{SA-tree} (\ttt{satree})  \cite{navarro2002searching}}   \\
\cmidrule(l){1-2} 
                   & No parameters \\
\cmidrule(l){1-2} 
\multicolumn{2}{c}{\textbf{bbtree} (\ttt{bbtree})  \cite{Cayton2008}}   \\
\cmidrule(l){1-2} 
                   & Common parameters: \ttt{bucketSize}, \ttt{chunkBucket}, and \ttt{maxLeavesToVisit} \\
\bottomrule
\multicolumn{2}{l}{\textbf{Note:} mnemonic method names are given in round brackets.}
\end{tabular}
\vspace{2em}
\end{table}

\subsubsection{GH-Tree}
A GH-tree \cite{Uhlmann:1991} is a binary tree. In each node the data set is divided using
two randomly selected pivots. Elements closer to one pivot are placed into a left
subtree, while elements closer to the second pivot are placed into a right subtree.

The following is an example of testing the GH-tree with the benchmarking utility \ttt{experiment}:
{
\footnotesize
\begin{verbatim}
release/experiment \
  --distType float --spaceType l2 --testSetQty 5 --maxNumQuery 100 \
  --knn 1 --range 0.1 \
  --dataFile ../sample_data/final8_10K.txt --outFilePrefix result \
  --method ghtree \
    --createIndex bucketSize=10,chunkBucket=1 \
    --queryTimeParams maxLeavesToVisit=10
\end{verbatim}
}

Our implementation of the GH-tree permits to answer queries both exactly and approximately (by specifying
the parameter \ttt{maxLeavesToVisit}). Yet, this implementation should be used only with metric spaces.

\subsubsection{List of Clusters}\label{SectionListClust}
The list of clusters \cite{chavez2005compact} is an exact search method for metric spaces,
which relies on flat (i.e., non-hierarchical) clustering.
Clusters are created sequentially starting by randomly selecting the first cluster center.
Then, close points are assigned to the cluster and the clustering procedure
is applied to the remaining points.
Closeness is defined either in terms of the maximum \ttt{radius},
or in terms of the maximum number (\ttt{bucketSize}) of points closest to the center.

Next we select cluster centers according to one of the policies:
random selection, a point closest to the previous center,
a point farthest from the previous center, a point that minimizes
the sum of distances to the previous center, and a point that maximizes  
the sum of distances to the previous center.
In our experience, a random selection strategy (a default one)
works well in most cases.

The search algorithm iterates over the constructed list of clusters and
checks if answers can potentially belong to the currently selected cluster
 (using the triangle inequality).
If the cluster can contain an answer,
each cluster element is compared directly against the query.
Next, we use the triangle inequality to verify if answers can
be outside the current cluster.
If this is not possible, the search is terminated.

We modified this exact algorithm by introducing an early termination condition.
The clusters are visited in the order of increasing distance from
the query to a cluster center.
The search process stops after vising a \ttt{maxLeavesToVisit} clusters.
Our version is supposed to work for metric spaces (and symmetric distance functions),
but it can also be used with mildly-nonmetric symmetric distances such as the cosine distance.

\newpage 

An example of testing the list of clusters using the \ttt{bucketSize} as a parameter to define
the size of the cluster:
{
\footnotesize
\begin{verbatim}
release/experiment \
  --distType float --spaceType l2 --testSetQty 5 --maxNumQuery 100 \
  --knn 1 --range 0.1 \
  --dataFile ../sample_data/final8_10K.txt --outFilePrefix result \
  --method list_clusters \
    --createIndex useBucketSize=1,bucketSize=100,strategy=random \
    --queryTimeParams maxLeavesToVisit=5
\end{verbatim}
}
An example of testing the list of clusters using the \ttt{radius} as a parameter to define
the size of the cluster:
{
\footnotesize
\begin{verbatim}
release/experiment \
  --distType float --spaceType l2 --testSetQty 5 --maxNumQuery 100 \
  --knn 1 --range 0.1 \
  --dataFile ../sample_data/final8_10K.txt --outFilePrefix result \
  --method list_clusters \
    --createIndex useBucketSize=0,radius=0.2,strategy=random \
    --queryTimeParams maxLeavesToVisit=5
\end{verbatim}
}

\subsubsection{SA-tree}
The Spatial Approximation tree (SA-tree)  \cite{navarro2002searching} aims
to approximate the Voronoi partitioning.
A data set is recursively divided by selecting several cluster centers in a greedy fashion.
Then, all remaining data points are assigned to the closest cluster center.

A cluster-selection procedure first randomly chooses the main center point and arranges the
remaining objects in the order of increasing distances to this center.
It then iteratively fills the set of clusters as follows: We start from the empty cluster 
list. Then, we iterate over the set of data points and check if there is a cluster center that
is closer to this point than the main center point. 
If no such cluster exists (i.e., the point is closer to the main center point than to any
of the already selected cluster centers), the point becomes a new cluster center 
(and is added to the list of clusters).
Otherwise, the point is added to the nearest cluster from the list.

After the cluster centers are selected, each of them is indexed recursively using the already described
algorithm. Before this, however, we check  if there are points that need to be reassigned to a different cluster.
Indeed, because the list of clusters keeps growing, we may miss the nearest cluster not yet
added to the list. To fix this, we need  to compute distances among every cluster point
and cluster centers that were not selected at the moment of the point's assignment to the cluster.

Currently, the SA-tree is an exact search method for metric spaces without any parameters.
The following is an example of testing the SA-tree with the benchmarking utility \ttt{experiment}:
{
\footnotesize
\begin{verbatim}
release/experiment \
  --distType float --spaceType l2 --testSetQty 5 --maxNumQuery 100 \
  --knn 1 --range 0.1 \
  --dataFile ../sample_data/final8_10K.txt --outFilePrefix result \
  --method satree 
\end{verbatim}
}

\subsubsection{bbtree}
A Bregman ball tree (bbtree) is an exact search method for Bregman divergences \cite{Cayton2008}.
The bbtree divides data into two clusters (each covered by a Bregman ball)
and recursively repeats this procedure for each cluster until the number of data points
in a cluster falls below \ttt{bucketSize}. Then, such clusters are stored as a single bucket.

At search time, the method relies on properties of Bregman divergences 
to compute the shortest distance to a covering ball. 
This is a rather expensive iterative procedure that 
may require several computations of direct and inverse gradients,
as well as of several distances.

Additionally, Cayton  \cite{Cayton2008} employed an early termination method:
The algorithm  can be told to stop after processing
a \ttt{maxLeavesToVisit} buckets.
The resulting method is an approximate search procedure.

Our implementation of the bbtree uses the same code to carry
out the nearest-neighbor and
the range searching. 
Such an implementation of the range searching is somewhat suboptimal
and a better approach exists \cite{cayton2009efficient}.


The following is an example of testing the bbtree with the benchmarking utility \ttt{experiment}:
{
\footnotesize
\begin{verbatim}
release/experiment \
  --distType float --spaceType kldivgenfast \
  --testSetQty 5 --maxNumQuery 100 \
  --knn 1 --range 0.1 \
  --dataFile ../sample_data/final8_10K.txt --outFilePrefix result \
  --method bbtree \
    --createIndex bucketSize=10  \
    --queryTimeParams maxLeavesToVisit=20 
\end{verbatim}
}

\subsection{Locality-sensitive Hashing Methods} \label{SectionLSH}
Locality Sensitive Hashing (LSH) \cite{indyk1998approximate,Kushilevitz_et_al:1998} is a class of methods employing hash functions that tend to have the same hash values for close points and different hash values for distant points. It is a probabilistic method in which the probability of having the same hash value is a monotonically decreasing function of the distance between two points (that we compare). A hash function that possesses this property is called \emph{locality sensitive}. 

Our library embeds the LSHKIT which provides locality sensitive hash functions in $L_1$ and $L_2$.
It supports only the nearest-neighbor (but not the range) search.
Parameters of LSH methods are summarized in Table~\ref{TableLSHParams}.
The LSH methods are not available under Windows.

\begin{table}[t!]
\caption{Parameters of LSH methods\label{TableLSHParams}}
\centering
\begin{tabular}{l@{\hspace{2mm}}p{3.5in}}
\toprule
\multicolumn{2}{c}{\textbf{Common parameters}}\\
\cmidrule(l){1-2} 
\ttt{W}  & A width of the window \cite{dong2011high}.  \\
\ttt{M}  & A number of atomic (binary hash functions),
which are concatenated to produce an integer hash value.  \\
\ttt{H}  & A size of the hash table.  \\
\ttt{L}  & The number hash tables.  \\

\cmidrule(l){1-2} 
\multicolumn{2}{c}{
\textbf{Multiprobe LSH: only for $L_2$} (\ttt{lsh\_multiprobe}) \cite{lv2007multi,Dong_et_al:2008,dong2011high} } \\
\cmidrule(l){1-2} 
                   & Common parameters: \ttt{W}, \ttt{M}, \ttt{H}, and \ttt{L} \\
\ttt{T}            & a number of probes \\
\ttt{desiredRecall}& a desired recall \\
\ttt{numSamplePairs}& a number of samples (P in lshkit)\\
\ttt{numSampleQueries}& a number of sample queries (Q in lshkit)\\
\ttt{tuneK}        & find optimal parameter for \knn, search
                     where $k$ is defined by this parameter \\
\cmidrule(l){1-2} 
\multicolumn{2}{c}{\textbf{LSH Gaussian: only for $L_2$  (\ttt{lsh\_gaussian}) } \cite{charikar2002similarity}}\\
\cmidrule(l){1-2} 
                   & Common parameters: \ttt{W}, \ttt{M}, \ttt{H}, and \ttt{L} \\
\cmidrule(l){1-2} 
\multicolumn{2}{c}{\textbf{LSH Cauchy: only for $L_1$ } (\ttt{lsh\_cauchy}) \cite{charikar2002similarity}} \\
\cmidrule(l){1-2} 
                   & Common parameters: \ttt{W}, \ttt{M}, \ttt{H}, and \ttt{L} \\
\cmidrule(l){1-2} 
\multicolumn{2}{c}{\textbf{LSH thresholding: only for $L_1$ } (\ttt{lsh\_threshold})  \cite{wang2007sizing,lv2004image}} \\
\cmidrule(l){1-2} 
                   & Common parameters: \ttt{M}, \ttt{H}, and \ttt{L} (\ttt{W} is not used)\\
\bottomrule
\multicolumn{2}{l}{\textbf{Note:} mnemonic method names are given in round brackets.}
\end{tabular}
\vspace{2em}
\end{table}

Random projections is a common approach to design locality sensitive hash functions.
These functions are composed from \texttt{M} binary hash functions $h_i(x)$.
A concatenation of the binary hash function values, i.e.,
 $h_1(x) h_2(x) \ldots h_M(x)$, is interpreted as a binary representation of the hash
function value $h(x)$. 
Pointers to objects with equal hash values (modulo \texttt{H})
are stored  in same cells of the hash table (of the size \texttt{H}).
If we used only one hash table, the probability of collision for two similar objects
would be too low. 
To increase the probability of finding a similar object multiple hash tables are used.
In that, we use a separate (randomly selected) hash function for each hash table.

To generate binary hash functions we first select a parameter \texttt{W} (called a \emph{width}).
Next, for every binary hash function, we draw a value $a_i$ from a $p$-stable distribution \cite{datar2004locality},
and a value $b_i$ from the uniform distribution with the support $[0, W]$.
Finally, we define $h_i(x)$ as:
$$
h_i(x) = \left\lfloor \frac{ a_i \cdot x + b_i  }{W}\right\rfloor,
$$
where $\lfloor x \rfloor$  is the \texttt{floor} function
and $x \cdot y$ denotes the scalar product of $x$ and $y$.

For the $L_2$ a standard Guassian distribution is $p$-stable, while for $L_1$ distance one can generate hash functions using a Cauchy distribution \cite{datar2004locality}.
For $L_1$, the LSHKIT defines another (``thresholding'') approach based on sampling.
It is supposed to work best for data points enclosed in  a cube $[a,b]^d$.
We omit the description here and refer the reader to the papers that introduced this method \cite{wang2007sizing,lv2004image}.

One serious drawback of the LSH is that it is memory-greedy.
To reduce the number of hash tables while keeping the collision probability for similar objects
sufficiently high, it was proposed to ``multi-probe'' the same hash table more than once.
When we obtain the hash value $h(x)$, we check (i.e., probe) not only the contents
of the hash table cell $h(x) \mod H$, but also contents of cells 
whose binary codes are ``close'' to $h(x)$ (i.e, they may differ by a small number of bits).
The LSHKIT, which is embedded in our library, contains a state-of-the-art
implementation of the multi-probe LSH that can automatically select
optimal values for parameters \texttt{M} and \texttt{W} to achieve a desired recall (remaining
parameters still need to be chosen manually). 

The following is an example of testing the multi-probe LSH with the benchmarking utility \ttt{experiment}.
We aim to achieve the recall value 0.25 (parameter \ttt{desiredRecall}) 
for the 1-NN search (parameter \ttt{tuneK}):
{
\footnotesize
\begin{verbatim}
release/experiment \
  --distType float --spaceType l2 --testSetQty 5 --maxNumQuery 100 \
  --knn 1  \
  --dataFile ../sample_data/final8_10K.txt --outFilePrefix result \
  --method lsh_multiprobe \
    --createIndex desiredRecall=0.25,tuneK=1,\
                  T=5,L=25,H=16535
\end{verbatim}
}

The classic version of the LSH for $L_2$ can be tested as follows:
{
\footnotesize
\begin{verbatim}
release/experiment \
  --distType float --spaceType l2 --testSetQty 5 --maxNumQuery 100 \
  --knn 1  \
  --dataFile ../sample_data/final8_10K.txt --outFilePrefix result \
  --method lsh_gaussian \
    --createIndex W=2,L=5,M=40,H=16535
\end{verbatim}
}

There are two ways to use LSH for $L_1$. First, 
we can invoke the implementation based on the Cauchy distribution:
{
\footnotesize
\begin{verbatim}
release/experiment \
  --distType float --spaceType l1 --testSetQty 5 --maxNumQuery 100 \
  --knn 1  \
  --dataFile ../sample_data/final8_10K.txt --outFilePrefix result \
  --method lsh_cauchy \
    --createIndex W=2,L=5,M=10,H=16535
\end{verbatim}
}

Second, we can use $L_1$ implementation based on thresholding.
Note that it does not use the width parameter \texttt{W}:
{
\footnotesize
\begin{verbatim}
release/experiment \
  --distType float --spaceType l1 --testSetQty 5 --maxNumQuery 100 \
  --knn 1  \
  --dataFile ../sample_data/final8_10K.txt --outFilePrefix result \
  --method lsh_threshold  \
    --createIndex L=5,M=60,H=16535
\end{verbatim}
}

\subsection{Projection-based Filter-and-Refine Methods}\label{SectionProjMethod}
Projection-based filter-and-refine methods operate by mapping data and query points
to a low(er) dimensional space (a \emph{projection} space) with a simple, 
easy to compute, distance function.
The search procedure consists in generation of candidate entries by searching
in a low-dimensional projection space with subsequent
refinement, where candidate entries are directly compared against the query
using the original distance function.

The number of candidate records is an important method parameter,
which can be specified as a fraction of the total number of data base entries
(parameter \ttt{dbScanFrac}). 

Different projection-based methods arise depending on: 
the type of a projection, the type of the projection space, 
and on the type of the search algorithm for the projection space.
A type of the projection can be specified via a method's parameter \ttt{projType}.
A dimensionality of the projection space is specified via a method's parameter \ttt{projDim}.

We support four well-known types of projections:
\begin{itemize}
\item Classic random projections using random orthonormal vectors (mnemonic name \ttt{rand});
\item Fastmap (mnemonic name \ttt{fastmap});
\item Distances to random reference points/pivots (mnemonic name \ttt{randrefpt});
\item Based on permutations \ttt{perm};
\end{itemize}
All but the classic random projections are distance-based and
can be applied to an arbitrary space with the distance function.
Random projections can be applied only to vector spaces.
A more detailed description of projection approaches is given in \S~\ref{SectionProjDetails}

We provide two basic implementations to generate candidates.
One is based on brute-force searching in the projected space and another builds a VP-tree
over objects' projections.
In what follows, these methods are described in detail.

\subsubsection{Brute-force projection search.}\label{SectionProjBruteForce}
In the brute-force approach, we scan the list of projections and compute the distance
between the projected query and a projection of every data point.
Then, we sort all data points in the order of increasing distance to the projected query.
A fraction (defined by \ttt{dbScanFrac}) of data points is compared directly against the query.
Top candidates (most closest entries) are identified using either the priority queue
or incremental sorting (\cite{Chavez2008incsort}). 
Incremental sorting is a more efficient approach enabled by default.
The mnemonic code of this method is \ttt{proj\_incsort}.

A choice of the distance in the projected space is governed by the parameter \ttt{useCosine}.
If it set to 1, the cosine distance is used (this makes most sense if we use the cosine distance
in the original space). By default $\ttt{useCosine}=0$, which forces the use of $L_2$ in the projected space. 

The following is an example of testing the brute-force search of projections with the benchmarking utility \ttt{experiment}:
{
\footnotesize
\begin{verbatim}
release/experiment \
  --distType float --spaceType cosinesimil --testSetQty 5 --maxNumQuery 100 \
  --knn 1 --range 0.1 \
  --dataFile ../sample_data/final8_10K.txt --outFilePrefix result \
  --method proj_incsort \
    --createIndex projType=rand,projDim=4 \
    --queryTimeParams useCosine=1,dbScanFrac=0.01
\end{verbatim}
}

\subsubsection{Projection VP-tree.}\label{SectionProjVPTree}
To avoid exhaustive search in the space of projections, 
it is possible to index projected vectors using a VP-tree.
The method's mnemonic name is \ttt{proj\_vptree}.
In that, one needs to specify both the parameters of the VP-tree  (see \S~\ref{SectionVPtree})
and the projection parameters as in the case of brute-force searching of projections (see \S~\ref{SectionProjBruteForce}). 

The major difference from the brute-force search over projections is that, instead of choosing between $L_2$ and cosine distance as the distance
in the projected space, one uses a methods' parameter \ttt{projSpaceType} to specify an arbitrary one.
Similar to the regular VP-tree implementation,
optimal $\alpha_{left}$ and $\alpha_{right}$ are determined by the utility \ttt{tune\_vptree} via a grid search like procedure (\ttt{tune\_vptree.exe} on Windows). 

This method, unfortunately, tends to perform worse than the VP-tree applied to the original space. 
The only exception are spaces with high intrinsic (and, perhaps, representational) dimensionality
where VP-trees (even with an approximate search algorithm) are useless unless dimensionality is reduced substantially.
One example is Wikipedia tf-idf vectors see:
\newline
\url{https://github.com/nmslib/nmslib/tree/v\LibVersion/manual/datasets.md}.

The following is an example of testing the VP-tree over projections with the benchmarking utility \ttt{experiment}:
{
\footnotesize
\begin{verbatim}
release/experiment \
  --distType float --spaceType cosinesimil --testSetQty 5 --maxNumQuery 100 \
  --knn 1 --range 0.1 \
  --dataFile ../sample_data/final8_10K.txt --outFilePrefix result \
  --method proj_vptree  \
    --createIndex projType=rand,projDim=4,projSpaceType=cosinesimil \
    --queryTimeParams alphaLeft=2,alphaRight=2,dbScanFrac=0.01
\end{verbatim}
}

\subsubsection{\textbf{OMEDRANK}.}\label{SectionOmedrank}
In OMEDRANK \cite{Fagin2003} there is a small set of voting pivots,
each of which ranks data points based on a somewhat imperfect notion of the distance from points to the query (computed
by a classic random projection or a projection of some different kind).
While each individual ranking is imperfect,
a more accurate ranking can be achieved by rank aggregation. 
When such a consolidating ranking is found, the most highly ranked objects from this
\emph{aggregate} ranking can be used as answers to a nearest-neighbor query.
Finding the aggregate ranking is an NP-complete problem that Fagin~et~al.~\cite{Fagin2003} solve only heuristically.

Technically, during the index time, each point in the original space is projected into a (low)er dimensional
vector space. The dimensionality of the projection is defined using a method's parameter  \ttt{numPivot}
(note that this is different from other projection methods).
Then, for each dimension $i$ in the projected space, we sort data points in the order of increasing value of
the \mbox{$i$-th} element of its projection.

We also divide the index in chunks each accounting for at most \ttt{chunkIndexSize} data points.
The search algorithm processes one chunk at a time. The idea is to make a chunk sufficiently small
so that auxiliary data structures fit into L1 or L2  cache.

The retrieval algorithm uses \ttt{numPivot} pointers $low_i$ and \ttt{numPivot} pointers $high_i$ ($low_i \le high_i$),
The \mbox{$i$-th} pair of pointers ($low_i$, $high_i$) indicate a start and an end position in the \mbox{$i$-th} list.
For each data point, we allocate a zero-initialized counter.
We further create a projection of the query and use \ttt{numPivot} binary searches to find 
\ttt{numPivot} data points that have the closest \mbox{$i$-th} projection coordinates.
In each of the $i$ list, we make both $high_i$ and $low_i$ point to the found data entries.
In addition, for each data point found, we increase its counter.
Note that a single data point may appear the closest with respect to more than one projection coordinate!

After that, we run a series of iterations. In each iteration, we increase \ttt{numPivot} pointers $high_i$ and
decrease \ttt{numPivot} pointers $low_i$ (unless we reached the beginning or the end of a list).
For each data entry at which the pointer points, we increase the value of the counter.
Obviously, when we complete traversal of all \ttt{numPivot} lists, each counter will have the value \ttt{numPivot} (recall
that each data point appears exactly once in each of the lists).
Thus, sooner or later the value of a counter becomes equal to or larger than $\ttt{numPivot} \times \ttt{minFreq}$, where \ttt{minFreq}
is a method's parameter, e.g., 0.5.

The first point whose counter becomes equal to or larger than $\ttt{numPivot} \times \ttt{minFreq}$, becomes the first candidate
entry to be compared directly against the query. 
The next point whose counter matches the threshold value $\ttt{numPivot} \times \ttt{minFreq}$, 
becomes the second candidate and so on so forth.
The total number of candidate entries is defined by the parameter \ttt{dbScanFrac}.
Instead of all \ttt{numPivot} lists, it its possible to use only \ttt{numPivotSearch} lists that correspond
to the smallest absolute value of query's projection coordinates. In this case, the counter threshold is  $\ttt{numPivotSearch} \times \ttt{minFreq}$.
By default, $\ttt{numPivot}=\ttt{numPivotSearch}$.

Note that parameters \ttt{numPivotSearch} and \ttt{dbScanFrac}
were introduced by us, they were not employed in the original version of OMEDRANK.

\newpage

The following is an example of testing OMEDRANK with the benchmarking utility \ttt{experiment}:
{
\footnotesize
\begin{verbatim}
release/experiment \
  --distType float --spaceType cosinesimil --testSetQty 5 --maxNumQuery 100 \
  --knn 1 --range 0.1 \
  --dataFile ../sample_data/final8_10K.txt --outFilePrefix result \
    --method omedrank \
    --createIndex projType=rand,numPivot=8  \
    --queryTimeParams minFreq=0.5,dbScanFrac=0.02
\end{verbatim}
}

\begin{table}
\caption{Parameters of projection-based filter-and-refine methods\label{TableSpaceProjMethods}}
\centering
\begin{tabular}{l@{\hspace{2mm}}p{3.5in}}
\toprule
\multicolumn{2}{c}{\textbf{Common parameters}}\\
\cmidrule(l){1-2} 
\ttt{projType}      & A type of projection.    \\
\ttt{projDim}       & Dimensionality of projection vectors. \\
\ttt{intermDim}     & An intermediate dimensionality used to reduce dimensionality via the hashing trick (used only for sparse vector spaces). \\
\ttt{dbScanFrac}    & A number of candidate records obtained during the filtering step. \\
\multicolumn{2}{c}{\textbf{Brute-force Projection Search} (\ttt{proj\_incsort}) } 
\\
\cmidrule(l){1-2} 
                     & Common parameters: \ttt{projType}, \ttt{projDim}, \ttt{intermDim}, \ttt{dbScanFrac} \\
 \ttt{useCosine}    & If set to one, we use the cosine distance in the projected space. By default (value zero),
                      $L_2$ is used. \\
 \ttt{useQueue}    & If set to one, we use the priority queue instead of incremental sorting. By default is zero.\\
                      
\cmidrule(l){1-2} 
\multicolumn{2}{c}{\textbf{Projection VP-tree} (\ttt{proj\_vptree}) } 
\\
\cmidrule(l){1-2} 
                     & Common parameters: \ttt{projType}, \ttt{projDim}, \ttt{intermDim}, \ttt{dbScanFrac} \\
 \ttt{projSpaceType} & Type of the space of projections                 \\
 \ttt{bucketSize}    & A maximum number of elements in a bucket/leaf.    \\
 \ttt{chunkBucket}   & Indicates if bucket elements should be stored contiguously in memory (1 by default).  \\
 \ttt{maxLeavesToVisit}  & An early termination parameter equal to the maximum number of buckets (tree leaves) visited by a search algorithm (2147483647 by default). \\
 \ttt{alphaLeft}/\ttt{alphaRight}   & A stretching coefficient $\alpha_{left}$/$\alpha_{right}$ in Eq.~(\ref{EqDecFunc}) \\
 \ttt{expLeft}/\ttt{expRight} & The left/right exponent in Eq.~(\ref{EqDecFunc}) \\
\cmidrule(l){1-2} 
\multicolumn{2}{c}{\textbf{OMEDRANK} \cite{Fagin2003} (\ttt{omedrank}) } 
\\
\cmidrule(l){1-2} 
                     & Common parameters: \ttt{projType}, \ttt{intermDim}, \ttt{dbScanFrac} \\
\ttt{numPivot}       & Projection dimensionality \\
\ttt{numPivotSearch} & Number of data point lists to be used in search \\
\ttt{minFreq}        & The threshold for being considered a candidate entry: whenever
                       a point's counter becomes $\ge\ttt{numPivotSearch}\times\ttt{minFreq}$,
                       this point is compared directly to the query.  \\
\ttt{chunkIndexSize} & A number of documents in one index chunk.  \\
\bottomrule
\multicolumn{2}{l}{\textbf{Note:} mnemonic method names are given in round brackets.}
\end{tabular}
\vspace{2em}
\end{table}

\subsection{Permutation-based Filtering Methods} \label{SectionPermMethod}

Rather than relying on distance values directly,
we can assess similarity of objects based on their
relative distances to reference points (i.e., pivots).
For each data point $x$,
we can arrange pivots $\pi$ in the order of increasing distances from $x$ (for
simplicity we assume that there are no ties).
This arrangement is called a \emph{permutation}.
The permutation is essentially a pivot ranking. 
Technically, it is a vector whose \mbox{$i$-th}
element keeps an (ordinal) position of the \mbox{$i$-th} pivot (in the set of pivots
sorted by a distance from $x$).

Computation of the permutation
is a mapping from a source space, which may not have coordinates, to a target vector space with integer coordinates. 
In our library, the distance between permutations is defined as either $L_1$ or $L_2$.
Values of the distance in the source space often correlates well with the distance in the target space
of permutations.
This property is exploited in permutation methods.
An advantage of permutation methods is that they are not relying on metric properties of the original distance 
and can be successfully applied to non-metric spaces \cite{Boytsov_and_Bilegsaikhan:nips2013,naidan2015permutation}.

Note that there is no simple relationship between the distance in the target space
and the distance in the source space. In particular, the distance in the target space is neither a lower nor an upper bound
for the distance in the source space.
Thus, methods based on indexing permutations are filtering methods that allow us to obtain only approximate solutions.
In the first step, we retrieve a certain number of candidate points whose permutations are sufficiently close
to the permutation of the query vector.
For these candidate data points, we compute an actual distance to the query, using the original distance function.
For almost all implemented permutation methods, 
the number of candidate objects can be controlled by a parameter \ttt{dbScanFrac} or \ttt{minCandidate}.

Permutation methods differ in how they index and process permutations.
In the following subsections, we briefly review implemented variants.
Parameters of these methods are summarized in Tables~\ref{TablePermMethodParams}-\ref{TablePermMethodParamsCont}.

\subsubsection{Brute-force permutation search.}
In the brute-force approach, we scan the list of permutation methods and compute the distance
between the permutation of the query and a permutation of every data point.
Then, we sort all data points in the order of increasing distance to the query permutation
and a fraction (\ttt{dbScanFrac}) of data points is compared directly against the query.

In the current version of the library, the brute-force search over
regular permutations is a special case of the brute-force search
over projections (see \ref{SectionProjBruteForce}), where the projection type is \ttt{perm}. 
There is also an additional brute-force filtering method, which relies on the so-called binarized permutations.
It is described in~\ref{SectionPermBinary}.

\subsubsection{Permutation Prefix Index  (PP-Index).}
In a permutation prefix index (PP-index),
 permutation are stored in a prefix tree 
of limited depth \cite{Esuli:2012}. A parameter \ttt{prefixLength}
defines the depth.
The filtering phase aims to find \ttt{minCandidate} candidate data points.
To this end, it first retrieves the data points whose prefix of the inverse pivot ranking is exactly the same
as that of the query. If we do not get enough candidate objects, we shorten the prefix
and repeat the procedure until we get a sufficient number of candidate entries.
Note that we do not the use the parameter \ttt{dbScanFrac} here.


The following is an example of testing the PP-index with the benchmarking utility \ttt{experiment}.
{
\footnotesize
\begin{verbatim}
release/experiment \
  --distType float --spaceType l2 --testSetQty 5 --maxNumQuery 100 \
  --knn 1 --range 0.1 \
  --dataFile ../sample_data/final8_10K.txt --outFilePrefix result \
  --method pp-index \
    --createIndex numPivot=4 \
    --queryTimeParams prefixLength=4,minCandidate=100
\end{verbatim}
}

\subsubsection{VP-tree index over permutations.}
We can use a VP-tree to index permutations.
This approach is similar to that of Figueroa and Fredriksson \cite{figueroa2009speeding}. 
We, however, rely on the approximate version of the VP-tree described in \S~\ref{SectionVPtree},
while Figueroa and Fredriksson use an exact one.
The ``sloppiness'' of the VP-tree search is governed by the stretching coefficients 
 \ttt{alphaLeft}   and \ttt{alphaRight} as well as by the exponents in Eq.~(\ref{EqDecFunc}).
In NMSLIB, the VP-tree index over permutations is a special
case of the projection VP-tree (see \S~\ref{SectionProjVPTree}).
There is also an additional VP-tree based method that indexes binarized permutations.
It is described in \S~\ref{SectionPermBinary}.

\subsubsection{Metric Inverted File}
(MI-File) relies on the inverted index over permutations \cite{amato2008approximate}.
We select (a potentially large) subset of pivots (parameter \ttt{numPivot}).
Using these pivots, we compute a permutation for every data point.
Then, \ttt{numPivotIndex} most closest pivots are memorized in a data file. 
If a pivot number $i$ is the \mbox{$pos$-th} most distant pivot for the object $x$,
we add the pair $(pos,x)$ to the posting list number $i$.
All posting lists are kept sorted in the order of the increasing first element 
(equal to the ordinal position of the pivot in a permutation).

During searching, we compute the permutation of the query and select
posting lists corresponding to \ttt{numPivotSearch} most closest pivots.
These posting lists are processed as follows: Imagine that we 
selected posting list $i$ and the position of pivot $i$ in the permutation
of the query is $pos$. Then, 
using the posting list $i$,
we retrieve all candidate records
for which the position of the pivot $i$ in their respective permutations
is from 
$pos - \mbox{maxPosDiff}$ 
to
$pos + \mbox{maxPosDiff}$.
This allows us to update the estimate for the $L_1$ distance between retrieved
candidate records' permutations and the permutation of the query (see \cite{amato2008approximate}
for more details).

Finally, we select at most $\mbox{dbScanFrac} \cdot N$ objects ($N$ is the total
number of indexed objects) with the smallest
estimates for the $L_1$ between their permutations and the permutation of the query.
These objects are compared directly against the query.
The filtering step of the MI-file is expensive. Therefore,
this method is efficient only for computationally-intensive distances.


An example of testing this method using the utility \texttt{experiment} is as follows:
{
\footnotesize
\begin{verbatim}
release/experiment \
  --distType float --spaceType l2 --testSetQty 5 --maxNumQuery 100 \
  --knn 1 --range 0.1 \
  --dataFile ../sample_data/final8_10K.txt --outFilePrefix result \
  --method mi-file  \
    --createIndex numPivot=128,numPivotIndex=16 \
    --queryTimeParams numPivotSearch=4,dbScanFrac=0.01
\end{verbatim}
}

\subsubsection{Neighborhood APProximation Index (NAPP).}\label{SectionNAPP}
Recently it was proposed to index pivot neighborhoods: 
For each data point, we compute distances to \ttt{numPivot} points
and select \ttt{numPivotIndex} (typically, much smaller than \ttt{numPivot}) pivots 
that are closest to the data point.
Then, we associate these \ttt{numPivotIndex} closest pivots with the data point 
via an inverted file \cite{tellez2013succinct}.
One can hope that for similar points two pivot neighborhoods will have a non-zero intersection.

To exploit this observation, our implementation of the pivot neighborhood indexing method retrieves all points that 
share at least \ttt{numPivotSearch} nearest neighbor pivots (using an inverted file).
Then, these candidates points can be compared directly against the query,
which works well for cheap distances like $L_2$.

For computationally expensive distances, one can add an additional filtering step by
setting the parameter \ttt{useSort} to one.
If \ttt{useSort} is one, all candidate entries are additionally sorted by the number of shared pivots
(in the decreasing order).
Afterwards, a subset of candidates are compared directly against the query. 
The size of the subset is defined by the parameter \ttt{dbScanFrac}.
When selecting the subset, we give priority to candidates sharing more common pivots with the query. 
This secondary filtering may eliminate less promising entries, but it incurs additional
computational costs, which may outweigh the benefits of additional filtering ``power'', if the distance is cheap.

In many cases, good performance can be achieved by selecting pivots randomly. However, we find
that pivots can also be engineered (more information on this topic will be published soon). 
To load external pivots, the user should specify
an index-time parameter \ttt{pivotFile}. The pivots should be in the same format as the data points.

Note that our implementation is different from that of Tellez~\cite{tellez2013succinct} in several ways.
First, we do not use a succinct inverted index. Second, we use a simple posting merging algorithm
based on counting (a \emph{ScanCount} algorithm}). 
Before a query is processed, we zero-initialize an array that keeps one
counter for every data point. As we traverse a posting list and encounter an entry corresponding to object
$i$, we increment a counter number $i$.
The ScanCount is known to be quite efficient \cite{li2008efficient}.

We also divide the index in chunks each accounting for at most \ttt{chunkIndexSize} data points.
The search algorithm processes one chunk at a time. The idea is to make a chunk sufficiently small
so that counters fit into L1 or L2  cache.

An example of running NAPP without the additional filtering stage:
{
\footnotesize
\begin{verbatim}
release/experiment \
  --distType float --spaceType l2 --testSetQty 5 --maxNumQuery 100 \
  --knn 1 --range 0.1 \
  --dataFile ../sample_data/final8_10K.txt --outFilePrefix result \
  --cachePrefixGS napp_gold_standard \
  --method napp \
    --createIndex numPivot=32,numPivotIndex=8,chunkIndexSize=1024 \
    --queryTimeParams numPivotSearch=8 \
    --saveIndex napp_index
\end{verbatim}
}
Note that NAPP is capable of saving/loading the meta index. However, in the bootstrapping
mode this is only possible if gold standard data is cached (hence, the option \ttt{--cachePrefixGS}).

An example of running NAPP \textbf{with} the additional filtering stage:
{
\footnotesize
\begin{verbatim}
release/experiment \
  --distType float --spaceType l2 --testSetQty 5 --maxNumQuery 100 \
  --knn 1 --range 0.1 \
  --dataFile ../sample_data/final8_10K.txt --outFilePrefix result \
  --method napp   \
    --createIndex numPivot=32,numPivotIndex=8,chunkIndexSize=1024 \
    --queryTimeParams useSort=1,dbScanFrac=0.01,numPivotSearch=8
\end{verbatim}
}

\subsubsection{Binarized permutation methods.}\label{SectionPermBinary}
Instead of computing the $L_2$ distance between two permutations,
we can binarize permutations and compute the Hamming distance between
binarized permutations. 
To this end, we select an adhoc binarization threshold \texttt{binThreshold} (the
number of pivots divided by two is usually a good setting).
All integer values smaller than \texttt{binThreshold} become zeros,
and values larger than or equal to \texttt{binThreshold} become ones.

The Hamming distance between binarized permutations can be computed much faster than $L_2$ or $L_1$ (see Table~\ref{TableSpaces}). This comes at a cost though, as the Hamming distance appears to be a worse proxy for the original distance than $L_2$ or $L_1$ (for the same
number of pivots).
One can compensate in quality by using more pivots. In our experiments,
it was usually sufficient to double the number of pivots.

The binarized permutation can be searched sequentially. 
An example of testing such a method using the utility \texttt{experiment} is as follows:
{
\footnotesize
\begin{verbatim}
release/experiment \
  --distType float --spaceType l2 --testSetQty 5 --maxNumQuery 100 \
  --knn 1 --range 0.1 \
  --dataFile ../sample_data/final8_10K.txt --outFilePrefix result \
  --method perm_incsort_bin \
    --createIndex numPivot=32,binThreshold=16 \
    --queryTimeParams dbScanFrac=0.05
\end{verbatim}
}

Alternatively, binarized permutations can be indexed using the VP-tree.
This approach is usually more efficient than searching binarized permutations sequentially,
 but one needs to tune additional parameters.
An example of testing such a method using the utility \texttt{experiment} is as follows:
{
\footnotesize
\begin{verbatim}
release/experiment \
  --distType float --spaceType l2 --testSetQty 5 --maxNumQuery 100 \
  --knn 1 --range 0.1 \
  --dataFile ../sample_data/final8_10K.txt --outFilePrefix result \
  --method perm_bin_vptree  \
    --createIndex numPivot=32 \
    --queryTimeParams alphaLeft=2,alphaRight=2,dbScanFrac=0.05
\end{verbatim}
}

\begin{table}
\caption{Parameters of permutation-based filtering methods\label{TablePermMethodParams}}
\centering
\begin{tabular}{l@{\hspace{2mm}}p{3.5in}}
\toprule
\multicolumn{2}{c}{\textbf{Common parameters}}\\
\cmidrule(l){1-2} 
\ttt{numPivot}      & A number of pivots. \\
\ttt{dbScanFrac}    & A number of candidate records obtained during the filtering step.
It is specified as a \emph{fraction} (not a percentage!) of 
the total number of data points in the data set. \\
\ttt{binThreshold}  & Binarization threshold.
If a value of an original permutation vector is below this threshold, 
it becomes 0 in the binarized permutation. If the 
value is above, the value is converted to 1.
\\
\cmidrule(l){1-2} 
\multicolumn{2}{c}{\textbf{Permutation Prefix Index} (\ttt{pp-index}) \cite{Esuli:2012}  }\\
\cmidrule(l){1-2} 
\ttt{numPivot}      & A number of pivots. \\
\ttt{minCandidate} & a minimum number of candidates to retrieve (note that we do not use
\ttt{dbScanFrac} here. \\
\ttt{prefixLength} & a maximum length of the tree prefix that is used to 
retrieve candidate records. \\
\ttt{chunkBucket} & 1 if we want to store vectors having the same permutation prefix
 in the same memory chunk (i.e., contiguously in memory) \\
\cmidrule(l){1-2} 
\multicolumn{2}{c}{\textbf{Metric Inverted File} (\ttt{mi-file}) \cite{amato2008approximate}  }\\
\cmidrule(l){1-2} 
                     & Common parameters: \ttt{numPivot} and \ttt{dbScanFrac}. \\
\ttt{numPivotIndex}  & a number of (closest) pivots to index                         \\
\ttt{numPivotSearch} & a number of (closest) pivots to use during searching          \\
\ttt{maxPosDiff}     & the maximum position difference permitted for searching      
in the inverted file \\
\cmidrule(l){1-2} 
\multicolumn{2}{c}{\textbf{Neighborhood Approximation Index} (\ttt{napp}) \cite{tellez2013succinct}  }\\
\cmidrule(l){1-2} 
                     & Common parameter \ttt{numPivot}. \\
\ttt{invProcAlg}     & An algorithm to merge posting lists. In practice, only \texttt{scan} worked  well. \\
\ttt{chunkIndexSize} & A number of documents in one index chunk.  \\
\ttt{indexThreadQty} & A number of indexhing threads. \\
\ttt{numPivotIndex}  & A number of closest pivots to be indexed. \\
\ttt{numPivotSearch} & A candidate entry should share this number of pivots with the query. 
This is a \textbf{query-time} parameter. \\
\bottomrule
\multicolumn{2}{l}{\textbf{Note:} mnemonic method names are given in round brackets.}
\end{tabular}
\end{table}

\begin{table}[t!]
\caption{Parameters of permutation-based filtering methods (continued) \label{TablePermMethodParamsCont}}
\centering
\begin{tabular}{l@{\hspace{2mm}}p{3.5in}}
\toprule
\multicolumn{2}{c}{\textbf{Brute-force search with incremental sorting for binarized permutations}}\\
\multicolumn{2}{c}{ (\ttt{perm\_incsort\_bin})  \cite{tellez2009brief} }\\
\cmidrule(l){1-2} 
                   & Common parameters: \ttt{numPivot}, \ttt{dbScanFrac}, \ttt{binThreshold}. \\
\cmidrule(l){1-2} 
\multicolumn{2}{c}{\textbf{VP-tree index over binarized permutations} (\ttt{perm\_bin\_vptree}) } \\
\multicolumn{2}{c}{ Similar to \cite{tellez2009brief}, but uses
an approximate search in the VP-tree. }  \\
\cmidrule(l){1-2} 
                   & Common parameters: \ttt{numPivot}, \ttt{dbScanFrac}, \ttt{binThreshold}. 
Note that \ttt{dbScanFrac} is a \textbf{query-time} parameter. \\
\cmidrule(l){1-2} 
 \ttt{alphaLeft}   & A stretching coefficient $\alpha_{left}$ in Eq.~(\ref{EqDecFunc}) \\
 \ttt{alphaRight}  & A stretching coefficient $\alpha_{right}$ in Eq.~(\ref{EqDecFunc}) \\
\bottomrule
\multicolumn{2}{l}{\textbf{Note:} mnemonic method names are given in round brackets.}
\end{tabular}
\end{table}

\subsection{Proximity/Neighborhood Graphs} \label{SectionProxGraph}
One efficient and effective search approach
relies on building a graph, where data points are graph nodes and edges connect sufficiently close points.
When edges connect nearest neighbor points, such graph is called a \knn graph (or a \href{https://en.wikipedia.org/wiki/Nearest_neighbor_graph}{nearest neighbor graph}).

In a proximity-graph a search process is a series of greedy sub-searches.
A sub-search starts at some, e.g., random node and proceeds to expanding the set of traversed nodes in a best-first fashion by following neighboring links. 
The algorithm resembles a Dijkstra's shortest-path algorithm in that, in each step, 
it selects an unvisited point closest to the query. 

There have been multiple stopping heuristics proposed.
For example, we can stop after visiting a certain number of nodes.
In NMSLIB, the sub-search terminates essentially when the candidate queue is exhausted.
Specifically, the candidate queue is expanded only with points that are closer to the query than the $k'$-\textit{th} closest point already discovered by the sub-search ($k'$ is a search parameter).
When we stop finding such points, the queue dwindles and eventually becomes empty. 
We also terminate the sub-search when the queue contains only points farther than the $k'$-\textit{th} closest point already discovered by the sub-search.
Note that the greedy search is only approximate and does not necessarily return all true nearest neighbors.

In our library we use several approaches to create proximity graphs, which are described below.
Parameters of these methods are summarized in Table~\ref{TableProxGraphs}.
Note that SW-graph and NN-descent have the parameter with the same name, namely, \ttt{NN}.
However, this parameter has a somewhat different interpretation depending on the method.
Also note that our proximity-graph methods support only the nearest-neighbor, but not the range search.

\subsubsection{Small World Graph (SW-graph).} \label{SectionSWGraph}
In the (Navigable) Small World graph (SW-graph),\footnote{SW-graph is also known as a Metrized Small-World (MSW) graph
and a Navigable Small World (NSW) graph.}
indexing is a bottom-up procedure that relies on the previously described greedy search algorithm.  
The number of restarts, though, is defined by a different parameter, i.e., \ttt{initIndexAttempts}.
We insert points one by one. For each data point, 
we find \ttt{NN} closest points using an already constructed index.
Then, we create an \emph{undirected} edge between a new graph node (representing a new point) 
and nodes that represent \ttt{NN} closest points found by the greedy search. 
Each sub-search starts from some, e.g., random node and proceeds expanding the candidate queue with points that are closer
than the \ttt{efConstruction}-\textit{th} closest point (\ttt{efConstruction} is an index-time parameter).
Similarly, the search procedure executes one or more sub-searches that start from some node.
The queue is expanded only with points that are closer than the \ttt{efSearch}-\textit{th} closest point.
The number of sub-searches is defined by the parameter \ttt{initSearchAttempts}.

Empirically, it was shown that this method often creates a navigable small world graph, where most nodes are separated by only a few edges (roughly logarithmic in terms of the overall number of objects) \cite{malkov2012scalable}.
A simpler and less efficient variant of this algorithm was presented at ICTA 2011 and SISAP 2012 \cite{Ponomarenko2011,malkov2012scalable}.
An improved variant appeared as an Information Systems publication \cite{malkov2014}.
In the latter paper, however, the values of \ttt{efSearch} and \ttt{efConstruction} are set equal to \ttt{NN}.
The idea of using values of \ttt{efSearch} and \ttt{efConstruction} potentially (much) larger than \ttt{NN}
was proposed by Malkov and Yashunin \cite{Malkov2016}.

The indexing algorithm is rather expensive and we accelerate it by running parallel searches in multiple threads. The number of threads is defined by the parameter \ttt{indexThreadQty}. By default, this parameter is equal to the number of virtual cores.
The graph updates are synchronized: If a thread  needs to add edges to a node or obtain
the list of node edges, it first locks a node-specific mutex.
Because different threads rarely update and/or access the same node simultaneously,
such synchronization creates little contention and, consequently,
our parallelization approach is efficient.
It is also necessary to synchronize updates for the list of graph nodes, 
but this operation takes little time compared to searching for \ttt{NN} neighboring points.

An example of testing this method using the utility \texttt{experiment} is as follows:
{
\footnotesize
\begin{verbatim}
release/experiment \
  --distType float --spaceType l2 --testSetQty 5 --maxNumQuery 100 \
  --knn 1  \
  --dataFile ../sample_data/final8_10K.txt --outFilePrefix result \
  --cachePrefixGS sw-graph \
  --method sw-graph \
    --createIndex NN=3,initIndexAttempts=5,indexThreadQty=4 \
    --queryTimeParams initSearchAttempts=1,efSearch=10 \
    --saveIndex sw-graph_index
\end{verbatim}
}
Note that SW-graph is capable of saving/loading the meta index. However, in the bootstrapping
mode this is only possible if gold standard data is cached (hence, the option \ttt{--cachePrefixGS}).

\subsubsection{Hierarchical Navigable Small World Graph (HNSW).} \label{SectionHNSW}

The Hierarchical Navigable Small World Graph (HNSW) \cite{Malkov2016} is a new search method,
a successor of the SW-graph.
HNSW can be much faster (especially during indexing) and is more robust.
However, the current implementation is still experimental and we will update it in the near future.

HNSW can be seen as a multi-layer and a multi-resolution variant of a proximity graph.
A ground (zero-level) layer includes all data points. The higher is the layer, the fewer points it has.
When a data point is added to HNSW, we select the maximum level $m$ randomly. 
In that, the probability of selecting level $m$ decreases exponentially with $m$.

Similarly to the SW-graph, the HNSW is constructed by inserting data points, one by one.
A new point is added to all layers starting from layer $m$ down to layer zero.
This is done using a search-based algorithm similar to that of the basic SW-graph. 
The quality is controlled by the parameter \ttt{efConstruction}.

Specifically, a search starts from the maximum-level layer and proceeds to lower layers by searching one layer at a time.
For all layers higher than the ground layer, the search algorithm is a 1-NN search 
that greedily follows the closest neighbor (this is equivalent to having \ttt{efConstruction}=1).
The closest point found at the layer $h+1$ is used as a starting point for the search carried out at the layer $h$. 
For the ground layer, we carry an \ttt{M}-NN search whose quality is controlled by the parameter \ttt{efConstruction}.
Note that the ground-layer search relies one the same algorithm as we use for the SW-graph (yet,
we use only a single sub-search, which is equivalent to setting \ttt{initIndexAttempts} to one).

An outcome of a search in a layer is a set of data points that are close to the new point.
Using one of the heuristics described by Malkov and Yashunin \cite{Malkov2016},
we select points from this set to become neighbors of the new point (in the layer's graph).
Note that unlike the older SW-graph, the new algorithm has a limit on the maximum number of neighbors.
If the limit is exceeded, the heuristics are used to keep only the best neighbors.
Specifically, the maximum number of neighbors in all layers but the ground layer is \ttt{maxM} (an index-time parameter,
which is equal to \ttt{M} by default).
The maximum number of neighbors for the ground layer is \ttt{maxM0} (an index-time parameter,
which is equal to $2\times$\ttt{M} by default). 
The choice of the heuristic is controlled by the parameter \ttt{delaunay\_type}.
Specifically, by default \ttt{delaunay\_type} is equal to 2. 
This default is generally quite good.
However, it maybe worth trying other viable options values: 0, 1, and 3.

A search algorithm is similar to the indexing algorithm. 
It starts from the maximum-level layer and proceeds to lower-level layers by searching one layer at a time.
For all layers higher than the ground layer, the search algorithm is a 1-NN search 
that greedily follows the closest neighbor (this is equivalent to having \ttt{efSearch}=1).
The closest point found at the layer $h+1$ is used as a starting point for the search carried out at the layer $h$. 
For the ground layer, we carry an \knn search whose quality is controlled by the parameter \ttt{efSearch} (in the 
paper by Malkov and Yashunin \cite{Malkov2016} this parameter is denoted as \ttt{ep}).
The ground-layer search relies one the same algorithm as we use for the SW-graph,
but it does not carry out multiple sub-searches starting from different random data points.

For $L_2$ and the cosine similarity, HNSW has optimized implementations, which are enabled by default.
To enforce the use of the generic algorithm, set the parameter \ttt{skip\_optimized\_index} to one.

Similar to SW-graph, the indexing algorithm can be expensive. 
It is, therefore, accelerated by running parallel searches in multiple threads. 
The number of threads is defined by the
parameter \ttt{indexThreadQty}. By default, this parameter is equal to the number
of virtual cores.

A sample command line to test HNSW using the utility \texttt{experiment}:
{
\footnotesize
\begin{verbatim}
release/experiment \
  --distType float --spaceType l2 --testSetQty 5 --maxNumQuery 100 \
  --knn 1  \
  --dataFile ../sample_data/final8_10K.txt --outFilePrefix result \
  --method hnsw \
    --createIndex M=10,efConstruction=20,indexThreadQty=4,searchMethod=0 \
    --queryTimeParams efSearch=10 
\end{verbatim}
}

HNSW is capable of saving an index for optimized $L_2$ and the cosine-similarity implementations. 
Here is an example for $L_2$:
{
\footnotesize
\begin{verbatim}
release/experiment \
  --distType float --spaceType cosinesimil --testSetQty 5 --maxNumQuery 100 \
  --knn 1  \
  --dataFile ../sample_data/final8_10K.txt --outFilePrefix result \
  --cachePrefixGS hnsw \
  --method hnsw \
    --createIndex M=10,efConstruction=20,indexThreadQty=4,searchMethod=4 \
    --queryTimeParams efSearch=10 
    --saveIndex hnsw_index
\end{verbatim}
}

\subsubsection{NN-Descent.} \label{SectionNNDescent}
The NN-descent is an iterative procedure initialized with randomly selected
nearest neighbors. In each iteration, a random sample of queries is selected
to participate in neighborhood propagation.

This process is governed by parameters \ttt{rho} and \ttt{delta}. 
Parameter \ttt{rho} defines a fraction of the data set that is randomly
sampled for neighborhood propagation. A good value that works
in many cases is $\ttt{rho}=0.5$. As the indexing algorithm iterates,
fewer and fewer neighborhoods change (when we attempt to improve the local
neighborhood structure via neighborhood propagation). 
The parameter \ttt{delta} defines a stopping condition in terms of a fraction
of modified edges in the \knnns graph (the exact definition can be inferred from code). 
A good default value is \ttt{delta}=0.001.
The indexing algorithm is multi-threaded: the method uses all available cores.

When NN-descent was incorporated into NMSLIB, 
there was no open-source search algorithm released, only the code to construct a \knnns graph.
Therefore, we use the same algorithm as for the SW-graph \cite{malkov2012scalable,malkov2014}.
The new, open-source, version of NN-descent (code-named \ttt{kgraph}), 
which does include the search algorithm,
can be found \href{https://github.com/aaalgo/kgraph}{on GitHub}.

Here is an example of testing this method using the utility \ttt{experiment}:
{
\footnotesize
\begin{verbatim}
release/experiment \
  --distType float --spaceType l2 --testSetQty 5 --maxNumQuery 100 \
  --knn 1  \
  --dataFile ../sample_data/final8_10K.txt --outFilePrefix result \
  --method nndes  \
    --createIndex NN=10,rho=0.5,delta=0.001 \
    --queryTimeParams initSearchAttempts=3
\end{verbatim}
}

\begin{table}
\caption{Parameters of proximity-graph based methods\label{TableProxGraphs}}
\centering
\begin{tabular}{l@{\hspace{2mm}}p{3.5in}}
\toprule
\multicolumn{2}{c}{\textbf{Common parameters}}\\
\cmidrule(l){1-2} 
\ttt{efSearch}            & The search depth: specifically, a sub-search is stopped,
                            when it cannot find a point closer than \ttt{efSearch}
                            points (seen so far) closest to the query.
\\
\cmidrule(l){1-2} 
\multicolumn{2}{c}{\textbf{SW-graph} (\ttt{sw-graph}) \cite{Ponomarenko2011,malkov2012scalable,malkov2014}  }\\
\cmidrule(l){1-2} 
\ttt{NN}                  & For a newly added point find this number of most closest points
                            that make the initial neighborhood of the point. When more points are added,
                            this neighborhood may be expanded. \\
\ttt{efConstruction}      & The depth of the search that is used to find neighbors during indexing.
                            This parameter is analogous to \ttt{efSearch}.\\
\ttt{initIndexAttempts}   & The number of random search restarts carried out to add one point.\\
\ttt{indexThreadQty}      & The number of indexing threads. The default value is
                            equal to the number of (logical) CPU cores. \\
\ttt{initSearchAttempts}  & A number of random search restarts. \\
\cmidrule(l){1-2} 
\multicolumn{2}{c}{\textbf{Hierarchical Navigable SW-graph} (\ttt{hnsw}) \cite{malkov2014}  }\\
\cmidrule(l){1-2} 
\ttt{mult}                & A scaling coefficient to determine the depth of a layered structure (see the paper by Malkov and Yashunin \cite{malkov2014} for details). A default value seems to be good enough. \\
\ttt{skip\_optimized\_index}   & Setting this parameter to one disables the use of the optimized implementations (for $L_2$
                                 and the cosine similarity).
                          \\
\ttt{maxM}                & The maximum number of neighbors in all layers but the ground layer (the default value seems to be good enough). \\
\ttt{maxM0}               & The maximum number of neighbors in the \emph{ground} layer (the default value seems to be good enough). \\
\ttt{M}                   & The size of the initial set of potential neighbors for the indexing phase. The set may be further 
                            pruned so that the overall number of neighbors does not exceed \ttt{maxM0} (for the ground layer) or \ttt{maxM}
                            (for all layers but the ground one).\\
 
\ttt{efConstruction}      & The depth of the search that is used to find neighbors during indexing (this parameter
                            is used only for the search in the ground layer). \\
\ttt{delaunay\_type}      & A type of the pruning heuristic: 0 indicates that we keep only \ttt{maxM} (or \ttt{maxM0} for the ground layer)
                            neighbors, 1 activates a heuristic described by Algorithm 4 \cite{malkov2014}  \\

\cmidrule(l){1-2} 
\multicolumn{2}{c}{\textbf{NN-descent} (\ttt{nndes}) \cite{dong2011efficient,malkov2012scalable,malkov2014}  }\\
\cmidrule(l){1-2} 
\ttt{NN}                  & For each point find this number of most closest points (neighbors). \\
\ttt{rho}                 & A fraction of the data set that is randomly sampled for neighborhood propagation.  \\
\ttt{delta}               & A stopping condition in terms of the fraction of updated edges in the \knnns graph. \\
\ttt{initSearchAttempts}  & A number of random search restarts. \\
\bottomrule
\multicolumn{2}{l}{\textbf{Note:} mnemonic method names are given in round brackets.}
\end{tabular}
\end{table}

\subsection{Miscellaneous Methods} \label{SectionMiscMeth}
Currently our major miscellaneous methods do not have parameters.  

\subsubsection{\textbf{Brute-force (sequential) searching}.}
To verify how the speed of brute-force searching
scales with the number of threads,
we provide a reference implementation of the sequential
searching.
For example, to benchmark sequential searching using two threads, 
one can type the following command:
{
\footnotesize
\begin{verbatim}
release/experiment \
  --distType float --spaceType l2 --testSetQty 5 --maxNumQuery 100 \
  --knn 1  \
  --dataFile ../sample_data/final8_10K.txt --outFilePrefix result \
  --method seq_search --threadTestQty 2
\end{verbatim}
}

\subsubsection{\textbf{Term-Based Uncompressed Inverted File}.}
For sparse data sets (especially when queries are much sparser than documents),
it can be useful to employ the inverted file.
The inverted file is a mapping from a set of dimensions to their respective posting lists. 
The posting list of a dimension is a sorted list of documents/vectors, where this dimension is non-zero. 
To answer queries, we use a classic  document-at-a-time (DAAT) processing algorithm \cite{DBLP:journals/ipm/Turtle1995}
with a priority queue (see Figure 5.3 in \cite{DBLP:books/daglib/0031897}),
which is a part of NMSLIB.
This algorithm is implemented in Lucene 6.0 (and earlier versions). 
However, our implementation is faster than Lucene  (1.5-2$\times$ faster on 
\emph{compr} and \emph{stack}) for several reasons:
our implementation does not use compression,
it does not explicitly compute BM25, 
it is written in C++ rather than Java

{
\footnotesize
\begin{verbatim}
release/experiment \
  --distType float --spaceType negdotprod_sparse_fast --testSetQty 5 \
  --maxNumQuery 100 \
  --knn 1  \
  --dataFile ../sample_data/sparse_wiki_5K.txt --outFilePrefix result \
  --method simple_invindx --threadTestQty 2
\end{verbatim}
}



\section{Notes on Efficiency}\label{SectionEfficiency}

\subsection{Efficiency of Distance Functions}
Note that improvement in efficiency and in the number of distance computations
obtained with slow distance functions can be overly optimistic.
That is, when a slow distance function is replaced with a more efficient version,
the improvements over sequential search may become far less impressive.
In some cases, the search method can become even slower than the brute-force
comparison against every data point.
This is why we believe that optimizing  computation of a distance function 
is equally important (and sometimes even more important) 
than designing better search methods.

In this library, we optimized several distance functions, 
especially non-metric functions that involve computations of logarithms.
An order of magnitude improvement can be achieved by pre-computing 
logarithms at index time and by approximating those logarithms that are not possible
to pre-compute (see \S~\ref{SectionJS} and \S~\ref{SectionBregman} for more details).
Yet, this doubles the size of an index.

The Intel compiler has a powerful math library, 
which allows one to efficiently compute several hard distance functions
such as the KL-divergence, the Jensen-Shanon divergence/metric, and
the $L_p$ spaces for non-integer values of $p$ more efficiently than in the case of GNU C++
and Clang.
In the Visual Studio's fast math mode (which is enabled in the provided project files)
it is also possible to compute some hard distances several times faster compared to GNU C++ and Clang.
Yet, our custom implementations are often much faster.
For example, in the case of the Intel compiler,
the custom implementation of the KL-divergence is 10 times faster 
than the standard one while
the custom implementation of the JS-divergence is two times faster.
In the case of the Visual studio, the custom KL-divergence is 
7 times as fast as the standard one, while the custom JS-divergence is 10 times faster.
Therefore,
doubling the size of the data set by storing pre-computed logarithms seems to be worthwhile.

Efficient implementations of some other distance functions 
rely on SIMD instructions. 
These instructions, available on most modern Intel and AMD processors, 
operate on small vectors. 
Some C++ implementations can be efficiently vectorized by both the GNU and Intel compilers.
That is, instead of the scalar operations the compiler would generate
more efficient SIMD instructions.
Yet, the code is not always vectorized, e.g., by the Clang.
And even the Intel compiler, fails to efficiently vectorize 
computation of the KL-divergence (with pre-computed logarithms).

There are also situations when efficient automatic vectorization
is hardly possible. For instance,
we provide an efficient implementation of the scalar product
for sparse \emph{single-precision} floating point vectors.
It relies on the all-against-all comparison SIMD instruction \texttt{\_mm\_cmpistrm}. 
However, it requires keeping the data in a special format,
which makes automatic vectorization impossible.

Intel SSE extensions that provide SIMD instructions are automatically detected
by all compilers but the Visual Studio.
If some SSE extensions are not available, the compilation process will produce warnings like the
following one:
{
\footnotesize
\begin{verbatim}
LInfNormSIMD: SSE2 is not available, defaulting to pure C++ implementation!
\end{verbatim}
}

\subsection{Cache-friendly Data Layout}
In our previous report \cite{Boytsov_and_Bilegsaikhan:sisap2013},
we underestimated a cost of a random memory access.
A more careful analysis showed that, 
on a recent laptop (Core i7, DDR3), 
a truly random access \href{http://searchivarius.org/blog/main-memory-similar-hard-drive}{``costs'' about 200 CPU cycles},
which may be 2-3 times longer than a computation of a cheap distance such as $L_2$.

Many implemented methods use some form of bucketing.
For example, in the VP-tree or bbtree we recursively decompose the space
until a partition contains at most \ttt{bucketSize} elements.
The buckets are searched sequentially,
which could be done much faster, if bucket objects were stored
in contiguous memory regions.
Thus, to check elements in a bucket we would need only one random memory access.

A number of methods support this optimized storage model.
It is activated by setting a parameter \ttt{chunkBucket} to 1.
If \ttt{chunkBucket} is set to 1, indexing is carried out in two stages.
At the first stage, a method creates unoptimized buckets,
each of which is an array of pointers to data objects.
Thus, objects are not necessarily contiguous in memory.
In the second stage, the method iterates over buckets,
allocates a contiguous chunk of memory,
which is sufficiently large to keep all bucket objects,
and copies bucket objects to this new chunk.

\textbf{Important note:}
Note that currently we do not delete old objects and do not deallocate the memory 
they occupy. Thus, 
if 
\ttt{chunkBucket} is set to 1,
the memory usage is overestimated.
In the future, we plan to address this issue.


\begin{thebibliography}{10}

\bibitem{amato2008approximate}
G.~Amato and P.~Savino.
\newblock Approximate similarity search in metric spaces using inverted files.
\newblock In {\em Proceedings of the 3rd international conference on Scalable
  information systems}, page~28. ICST (Institute for Computer Sciences,
  Social-Informatics and Telecommunications Engineering), 2008.

\bibitem{Beecks:2013}
C.~Beecks.
\newblock {\em Distance based similarity models for content based multimedia
  retrieval}.
\newblock PhD thesis, 2013.

\bibitem{Beecks:2010}
C.~Beecks, M.~S. Uysal, and T.~Seidl.
\newblock Signature quadratic form distance.
\newblock In {\em Proceedings of the ACM International Conference on Image and
  Video Retrieval}, CIVR '10, pages 438--445, New York, NY, USA, 2010. ACM.

\bibitem{bingham2001random}
E.~Bingham and H.~Mannila.
\newblock Random projection in dimensionality reduction: applications to image
  and text data.
\newblock In {\em Proceedings of the seventh ACM SIGKDD international
  conference on Knowledge discovery and data mining}, pages 245--250. ACM,
  2001.

\bibitem{Boytsov_and_Bilegsaikhan:sisap2013}
L.~Boytsov and B.~Naidan.
\newblock Engineering efficient and effective \mbox{Non-Metric Space Library}.
\newblock In N.~Brisaboa, O.~Pedreira, and P.~Zezula, editors, {\em Similarity
  Search and Applications}, volume 8199 of {\em Lecture Notes in Computer
  Science}, pages 280--293. Springer Berlin Heidelberg, 2013.

\bibitem{Boytsov_and_Bilegsaikhan:nips2013}
L.~Boytsov and B.~Naidan.
\newblock Learning to prune in metric and non-metric spaces.
\newblock In {\em Advances in Neural Information Processing Systems}, 2013.

\bibitem{bozkaya1999indexing}
T.~Bozkaya and M.~Ozsoyoglu.
\newblock Indexing large metric spaces for similarity search queries.
\newblock {\em ACM Transactions on Database Systems (TODS)}, 24(3):361--404,
  1999.

\bibitem{Bregman:1967}
L.~Bregman.
\newblock The relaxation method of finding the common point of convex sets and
  its application to the solution of problems in convex programming.
\newblock {\em \{USSR\} Computational Mathematics and Mathematical Physics},
  7(3):200 -- 217, 1967.

\bibitem{DBLP:books/daglib/0031897}
S.~B{\"{u}}ttcher, C.~L.~A. Clarke, and G.~V. Cormack.
\newblock {\em Information Retrieval - Implementing and Evaluating Search
  Engines}.
\newblock {MIT} Press, 2010.

\bibitem{Cayton2008}
L.~Cayton.
\newblock Fast nearest neighbor retrieval for bregman divergences.
\newblock In {\em Proceedings of the 25th international conference on Machine
  learning}, ICML '08, pages 112--119, New York, NY, USA, 2008. ACM.

\bibitem{cayton2009efficient}
L.~Cayton.
\newblock Efficient bregman range search.
\newblock In {\em Advances in Neural Information Processing Systems}, pages
  243--251, 2009.

\bibitem{charikar2002similarity}
M.~S. Charikar.
\newblock Similarity estimation techniques from rounding algorithms.
\newblock In {\em Proceedings of the thiry-fourth annual ACM symposium on
  Theory of computing}, pages 380--388. ACM, 2002.

\bibitem{chavez2005compact}
E.~Ch{\'a}vez and G.~Navarro.
\newblock A compact space decomposition for effective metric indexing.
\newblock {\em Pattern Recognition Letters}, 26(9):1363--1376, 2005.

\bibitem{Chavez_et_al:2001a}
E.~Ch\'{a}vez, G.~Navarro, R.~Baeza-Yates, and J.~L. Marroquin.
\newblock Searching in metric spaces.
\newblock {\em ACM Computing Surveys}, 33(3):273--321, 2001.

\bibitem{datar2004locality}
M.~Datar, N.~Immorlica, P.~Indyk, and V.~S. Mirrokni.
\newblock Locality-sensitive hashing scheme based on p-stable distributions.
\newblock In {\em Proceedings of the twentieth annual symposium on
  Computational geometry}, pages 253--262. ACM, 2004.

\bibitem{dong2011high}
W.~Dong.
\newblock {\em High-Dimensional Similarity Search for Large Datasets}.
\newblock PhD thesis, Princeton University, 2011.

\bibitem{dong2011efficient}
W.~Dong, C.~Moses, and K.~Li.
\newblock Efficient k-nearest neighbor graph construction for generic
  similarity measures.
\newblock In {\em Proceedings of the 20th international conference on World
  wide web}, pages 577--586. ACM, 2011.

\bibitem{Dong_et_al:2008}
W.~Dong, Z.~Wang, W.~Josephson, M.~Charikar, and K.~Li.
\newblock Modeling lsh for performance tuning.
\newblock In {\em Proceedings of the 17th ACM conference on Information and
  knowledge management}, CIKM '08, pages 669--678, New York, NY, USA, 2008.
  ACM.

\bibitem{endres2003new}
D.~M. Endres and J.~E. Schindelin.
\newblock A new metric for probability distributions.
\newblock {\em Information Theory, IEEE Transactions on}, 49(7):1858--1860,
  2003.

\bibitem{Esuli:2012}
A.~Esuli.
\newblock Use of permutation prefixes for efficient and scalable approximate
  similarity search.
\newblock {\em Inf. Process. Manage.}, 48(5):889--902, Sept. 2012.

\bibitem{Fagin2003}
R.~Fagin, R.~Kumar, and D.~Sivakumar.
\newblock Efficient similarity search and classification via rank aggregation.
\newblock In {\em Proceedings of the 2003 ACM SIGMOD International Conference
  on Management of Data}, SIGMOD '03, pages 301--312, New York, NY, USA, 2003.
  ACM.

\bibitem{faloutsos1995fastmap}
C.~Faloutsos and K.-I. Lin.
\newblock {\em FastMap: A fast algorithm for indexing, data-mining and
  visualization of traditional and multimedia datasets}, volume~24.
\newblock ACM, 1995.

\bibitem{figueroa2009speeding}
K.~Figueroa and K.~Frediksson.
\newblock Speeding up permutation based indexing with indexing.
\newblock In {\em Proceedings of the 2009 Second International Workshop on
  Similarity Search and Applications}, pages 107--114. IEEE Computer Society,
  2009.

\bibitem{Chavez2008incsort}
E.~Gonzalez, K.~Figueroa, and G.~Navarro.
\newblock Effective proximity retrieval by ordering permutations.
\newblock {\em Pattern Analysis and Machine Intelligence, IEEE Transactions
  on}, 30(9):1647--1658, 2008.

\bibitem{indyk1998approximate}
P.~Indyk and R.~Motwani.
\newblock Approximate nearest neighbors: towards removing the curse of
  dimensionality.
\newblock In {\em Proceedings of the thirtieth annual ACM symposium on Theory
  of computing}, pages 604--613. ACM, 1998.

\bibitem{Kushilevitz_et_al:1998}
E.~Kushilevitz, R.~Ostrovsky, and Y.~Rabani.
\newblock Efficient search for approximate nearest neighbor in high dimensional
  spaces.
\newblock In {\em Proceedings of the 30th annual ACM symposium on Theory of
  computing}, STOC '98, pages 614--623, New York, NY, USA, 1998. ACM.

\bibitem{Levenshtein:1966}
V.~Levenshtein.
\newblock Binary codes capable of correcting deletions, insertions, and
  reversals.
\newblock {\em Doklady Akademii Nauk SSSR,}, 163(4):845--848, 1966.

\bibitem{li2008efficient}
C.~Li, J.~Lu, and Y.~Lu.
\newblock Efficient merging and filtering algorithms for approximate string
  searches.
\newblock In {\em Data Engineering, 2008. ICDE 2008. IEEE 24th International
  Conference on}, pages 257--266. IEEE, 2008.

\bibitem{lv2004image}
Q.~Lv, M.~Charikar, and K.~Li.
\newblock Image similarity search with compact data structures.
\newblock In {\em Proceedings of the thirteenth ACM international conference on
  Information and knowledge management}, pages 208--217. ACM, 2004.

\bibitem{lv2007multi}
Q.~Lv, W.~Josephson, Z.~Wang, M.~Charikar, and K.~Li.
\newblock Multi-probe lsh: efficient indexing for high-dimensional similarity
  search.
\newblock In {\em Proceedings of the 33rd international conference on Very
  large data bases}, pages 950--961. VLDB Endowment, 2007.

\bibitem{malkov2012scalable}
Y.~Malkov, A.~Ponomarenko, A.~Logvinov, and V.~Krylov.
\newblock Scalable distributed algorithm for approximate nearest neighbor
  search problem in high dimensional general metric spaces.
\newblock In {\em Similarity Search and Applications}, pages 132--147.
  Springer, 2012.

\bibitem{malkov2014}
Y.~Malkov, A.~Ponomarenko, A.~Logvinov, and V.~Krylov.
\newblock Approximate nearest neighbor algorithm based on navigable small world
  graphs.
\newblock {\em Inf. Syst.}, 45:61--68, 2014.

\bibitem{Malkov2016}
Y.~A. {Malkov} and D.~A. {Yashunin}.
\newblock {Efficient and robust approximate nearest neighbor search using
  Hierarchical Navigable Small World graphs}.
\newblock {\em ArXiv e-prints}, Mar. 2016.

\bibitem{naidan2015permutation}
B.~Naidan, L.~Boytsov, and E.~Nyberg.
\newblock Permutation search methods are efficient, yet faster search is
  possible.
\newblock {\em {PVLDB}}, 8(12):1618--1629, 2015.

\bibitem{navarro2002searching}
G.~Navarro.
\newblock Searching in metric spaces by spatial approximation.
\newblock {\em The VLDB Journal}, 11(1):28--46, 2002.

\bibitem{Needleman_and_Wunsch:1970}
S.~B. Needleman and C.~D. Wunsch.
\newblock A general method applicable to the search for similarities in the
  amino acid sequence of two proteins.
\newblock {\em J Mol Biol}, 48(3):443--453, March 1970.

\bibitem{Ponomarenko2011}
A.~Ponomarenko, Y.~Malkov, A.~Logvinov, , and V.~Krylov.
\newblock Approximate nearest neighbor search small world approach, 2011.
\newblock Available at
  {\url{http://www.iiis.org/CDs2011/CD2011IDI/ICTA_2011/Abstract.asp?myurl=CT175ON.pdf}}.

\bibitem{ILSVRCarxiv14}
O.~Russakovsky, J.~Deng, H.~Su, J.~Krause, S.~Satheesh, S.~Ma, Z.~Huang,
  A.~Karpathy, A.~Khosla, M.~Bernstein, A.~C. Berg, and L.~Fei-Fei.
\newblock {ImageNet Large Scale Visual Recognition Challenge}, 2014.

\bibitem{Sakoe_and_Chiba:1971}
H.~Sakoe and S.~Chiba.
\newblock A dynamic programming approach to continuous speech recognition.
\newblock In {\em Proceedings of the Seventh International Congress on
  Acoustics}, pages 65--68, August 1971.
\newblock paper 20C13.

\bibitem{Sankoff:2000}
D.~Sankoff.
\newblock The early introduction of dynamic programming into computational
  biology.
\newblock {\em Bioinformatics}, 16(1):41--47, 2000.

\bibitem{schlegel2011fast}
B.~Schlegel, T.~Willhalm, and W.~Lehner.
\newblock Fast sorted-set intersection using simd instructions.
\newblock In {\em ADMS@ VLDB}, pages 1--8, 2011.

\bibitem{tellez2009brief}
E.~S. T{\'e}llez, E.~Ch{\'a}vez, and A.~Camarena-Ibarrola.
\newblock A brief index for proximity searching.
\newblock In {\em Progress in Pattern Recognition, Image Analysis, Computer
  Vision, and Applications}, pages 529--536. Springer, 2009.

\bibitem{tellez2013succinct}
E.~S. Tellez, E.~Ch{\'a}vez, and G.~Navarro.
\newblock Succinct nearest neighbor search.
\newblock {\em Information Systems}, 38(7):1019--1030, 2013.

\bibitem{DBLP:journals/ipm/Turtle1995}
H.~R. Turtle and J.~Flood.
\newblock Query evaluation: Strategies and optimizations.
\newblock {\em Inf. Process. Manage.}, 31(6):831--850, 1995.

\bibitem{Uhlmann:1991}
J.~Uhlmann.
\newblock Satisfying general proximity similarity queries with metric trees.
\newblock {\em Information Processing Letters}, 40:175--179, 1991.

\bibitem{Velichko_and_Zagoruyko:1970}
V.~Velichko and N.~Zagoruyko.
\newblock Automatic recognition of 200 words.
\newblock {\em International Journal of Man-Machine Studies}, 2(3):223 -- 234,
  1970.

\bibitem{Vintsyuk:1968}
T.~Vintsyuk.
\newblock Speech discrimination by dynamic programming.
\newblock {\em Cybernetics}, 4(1):52--57, 1968.

\bibitem{Wagner_and_Fischer:1974}
R.~A. Wagner and M.~J. Fischer.
\newblock The string-to-string correction problem.
\newblock {\em Journal of the ACM}, 21(1):168--173, 1974.

\bibitem{wang2007sizing}
Z.~Wang, W.~Dong, W.~Josephson, Q.~Lv, M.~Charikar, and K.~Li.
\newblock Sizing sketches: a rank-based analysis for similarity search.
\newblock {\em ACM SIGMETRICS Performance Evaluation Review}, 35(1):157--168,
  2007.

\bibitem{Yianilos:1993}
P.~N. Yianilos.
\newblock Data structures and algorithms for nearest neighbor search in general
  metric spaces.
\newblock In {\em Proceedings of the Fourth Annual ACM-SIAM Symposium on
  Discrete Algorithms}, SODA '93, pages 311--321, Philadelphia, PA, USA, 1993.
  Society for Industrial and Applied Mathematics.

\end{thebibliography}

\begin{appendix}
\section{Description of Projection Types}\label{SectionProjDetails}
\subsection{The classic random projections} 
The classic random projections work only for vector spaces (both sparse and dense).
At index time, we generate \ttt{projDim} vectors by sampling their elements from
the standard normal distribution $\mathcal{N}(0,1)$ and orthonormalizing them. \footnote{If 
the dimensionality of the projection space is larger than the dimensionality of the original
space, only the first \ttt{projDim} vectors are orthonormalized. The remaining are simply
divided by their norms.}
Coordinates in the projection spaces are obtained by computing scalar products
between a given vector and each of the \ttt{projDim} randomly generated vectors.

In the case of sparse vector spaces, the dimensionality is first reduced via the hashing trick:
the value of the element $i$ is equal to the sum of values for all elements whose indices are 
hashed into number $i$. 
After hashing, classic random projections are applied. 
The dimensionality of the intermediate space is defined by a method's parameter \ttt{intermDim}. 

The hashing trick is used purely for efficiency reasons. 
However, for large enough values of the intermediate
dimensionality, it has virtually no adverse affect on performance.
For example, in the case of Wikipedia tf-idf vectors (see  
\url{https://github.com/nmslib/nmslib/tree/v\LibVersion/manual/datasets.md}),
it is safe to use the value \ttt{intermDim}=4096.

Random projections work best if both the source and the target space are Euclidean,
whereas the distance is either $L_2$ or the cosine distance.
In this case, there are theoretical guarantees that the projection preserves
well distances in the original space (see e.g. \cite{bingham2001random}).

\subsection{FastMap} 
FastMap introduced by Faloutsos and Lin \cite{faloutsos1995fastmap}
is also a type of the random-projection method. 
At indexing time, we randomly select $projDim$ pairs $A_i$ and $B_i$.
The \mbox{$i$-\textit{th}} coordinate of vector $x$ is computed using the formula:
\begin{equation}\label{EqFastMap}
\mbox{\ttt{FastMap}}_{i}(x)  = \frac{d(A_i,x)^2 - d(B_i,x_i)^2 + d(A_i,B_i)}{2 d(A_i,B_i)^2}
\end{equation}
Given points $A$ and $B$ in the Euclidean space, Eq.~\ref{EqFastMap} gives the length of the
orthogonal projection of $x$ to the line connecting $A$ and $B$.
However, FastMap can be used in non-Euclidean spaces as well.

\subsection{Distances to the Random Reference Points} 
This method is a folklore projection approach,
where the \mbox{$i$-\textit{th}} coordinate of point $x$ in the projected space is computed as simply $d(x, \pi_i)$,
where $\pi_i$ is a pivot in the original space, i.e., a randomly selected reference point.
Pivots are selected once during indexing time.

\subsection{Permutation-based Projections.}
In this approach, we also select \ttt{projDim} pivots at index time.
However, instead of using raw distances to the pivots,
we rely on ordinal positions of pivots sorted by their distance to a point.  
A more detailed description is given in \S~\ref{SectionPermMethod}.
\end{appendix}

\end{document}